%% file: 00_main.tex
\documentclass[10pt,a4paper,onecolumn,accepted=2024-07-14]{quantumarticle}
\pdfoutput=1
\usepackage[utf8]{inputenc}
\usepackage[english]{babel}
\usepackage[T1]{fontenc}
\usepackage[numbers]{natbib}

\usepackage{auxi} 

\usepackage{geometry}

\usepackage{hyperref}
\hypersetup{
    colorlinks=true,
    linkcolor=blue,
    citecolor=blue,
    filecolor=magenta,      
    urlcolor=cyan,
    pdfpagemode=FullScreen,
    }
    
\usepackage{subcaption}

\renewcommand{\o}{&\hspace{-3mm}\otimes&\hspace{-3mm}}
\renewcommand{\N}{\mathbb{N}}
\renewcommand{\F}{\mathbb{F}}
\newcommand{\m}{\overline{m}}
\newcommand{\e}{\textup{e}}

\newcommand{\C}{\mathbb{C}}
\newcommand{\T}{{\!\mathsf{T}}}
\newcommand{\sprod}{\scalebox{.8}{$\prod$}}

\DeclareMathOperator{\rank}{rank}
\DeclareMathOperator{\Cay}{Cay}

\begin{document}
\title{Classical product code constructions for quantum Calderbank-Shor-Steane codes}
\date{}
\author{Dimiter Ostrev}
\affiliation{Institute of Communications and Navigation, German Aerospace Center (DLR), 82234 We{\ss}ling, Germany}
\author{Davide Orsucci}
\affiliation{Institute of Communications and Navigation, German Aerospace Center (DLR), 82234 We{\ss}ling, Germany}
\author{Francisco L\'azaro}
\affiliation{Institute of Communications and Navigation, German Aerospace Center (DLR), 82234 We{\ss}ling, Germany}
\author{Balazs Matuz}
\affiliation{Institute of Communications and Navigation, German Aerospace Center (DLR), 82234 We{\ss}ling, Germany}
\maketitle

\begin{abstract}
Several notions of code products are known in quantum error correction, such as hyper-graph products, homological products, lifted products, balanced products, to name a few. In this paper we introduce a new product code construction which is a natural generalisation of classical product codes to quantum codes: starting from a set of component Calderbank-Shor-Steane (CSS) codes, a larger CSS code is obtained where both $X$ parity checks and $Z$ parity checks are associated to classical product codes. We deduce several properties of product CSS codes from the properties of the component codes, including bounds to the code distance, and show that built-in redundancies in the parity checks result in so-called meta-checks which can be exploited to correct syndrome read-out errors. We then specialise to the case of single-parity-check (SPC) product codes which in the classical domain are a common choice for constructing product codes. Logical error rate simulations of a SPC $3$-fold product CSS code having parameters $[[512,174,8]]$ are shown under both a maximum likelihood decoder for the erasure channel and belief propagation decoding for depolarising noise. We compare the results with other codes of comparable length and dimension, including a code from the family of asymptotically good  Tanner codes. We observe that our reference product CSS code outperforms all the other examined codes. 
\end{abstract}

\section{Introduction}

Quantum information processing offers promising new solutions to problems in computation, communication, and cryptography~\cite{nielsen2010quantum}. The presence of noise in all quantum operations, however, results in the degradation of the quantum information present in a quantum system and is one of the major obstacles to the development and widespread application of quantum technologies. Calderbank-Shor-Steane (CSS) codes~\cite{calderbank1996good, steane1996error} and, more generally, stabilizer codes~\cite{gottesman1997stabilizer} can be employed to encode quantum information in a way that is resilient to errors affecting the physical qubits. Quantum error correction can then be applied to preserve the encoded quantum information in the presence of noise, with the ultimate goals of realising fully fault-tolerant quantum computation and robust encoders and decoders for quantum state transmission over noisy channels~\cite{hayashi2006quantum}. 

A quantum error correction approach that has gained prominence in the last few years is the use of quantum low-density parity-check (LDPC) codes~\cite{breuckmann2021quantum}, i.e.,  stabilizer codes having syndrome measurements of low weight. Low-weight stabilizers are considered favorable for fault-tolerance, since the quantum gates and measurements that implement them are also subject to noise.
In the quantum information community LDPC is reserved to codes that furthermore feature a high encoding rate (i.e., a high ratio of logical qubits to physical qubits). In fact, many long-established stabilizer code families, such as the well-known toric code, only encode a constant number of logical qubits and thus their encoding rate is inversely proportional to the number of employed physical qubits. Recent works have showcased that, for realistic physical error rates and practically useful blocks sizes, LDPC codes achieve error correction performances similar to that of the toric code but with encoding rates around ten times higher~\cite{bravyi2024highthreshold}. This demonstrates that the search for the best-performing stabilizer codes is far from concluded and that hitherto unexplored code families may still yield improvements for certain applications. 

In this work we introduce a novel natural method to construct LDPC-like CSS codes based on \textit{classical product codes}. The construction is very structured and yet flexible enough to produce codes exhibiting favourable properties. Our endeavour started from the observation that classical product codes were historically one of the first efficient coding constructions~\cite{elias1954error}. In general, classical code construction methods cannot be directly applied in the quantum domain: the commutativity of all stabilizer generators must be enforced, which imposes additional constraints on the code structure. Prior to this work, the only CSS code based on classical product codes was proposed by Hivadi in~\cite{hivadi2018quantum}; however, Hivadi's construction was restricted to the two-fold product of single parity check (SPC) codes, yielding codes with minimum distance only equal to 4.

\subsection{Comparison with other product code constructions}
In the classical domain, product operations produce powerful codes from two (or more) weaker component codes. For two component codes, the code words can be represented as a rectangular array whose columns are code words of one code and rows are code words of the other, see Figure~\ref{fig:ProductCode}. A further construction, \textit{tensor product codes}, was originally introduced in~\cite{wolf1965codes} with applications, e.g., to channels with burst errors. The code's parity-check matrix (PCM) is obtained by taking the tensor product of the PCMs of two codes. Tensor product codes were further generalized in~\cite{imai1981generalized} and extended to the construction of quantum error correcting codes in~\cite{fan2017quantumtensorproductcodes}. 

The first notion of product codes introduced in the quantum information literature has been that of \textit{hyper-graph product}~\cite{tillich2013quantum}, which gave the first family of quantum LDPC codes having constant rate and minimum distance scaling as the square root of the block length. Later, it was realised that the hyper-graph product can be seen as a special case of the \textit{homological product} of chain complexes~\cite{bravyi2014homological}, which can be leveraged, e.g., to construct quantum error correcting codes exhibiting single-shot error correction properties~\cite{quintavalle2021singleshot}. Asymptotically superior codes have been built using so-called \textit{lifted products}~\cite{panteleev2021degenerate} and \textit{balanced products}~\cite{breuckmann2021balanced} constructions. These techniques have culminated in the seminal result of constructing the first known families of asymptotically good quantum LDPC codes, i.e.\ having constant stabilizer weight, constant rate, and linearly growing minimum distance~\cite{panteleev2022asymptotically}. See also~\cite{breuckmann2021quantum} for more details on quantum product code constructions. 

These quantum product techniques are built to automatically satisfy the stabilizer commutativity condition and therefore are in spirit and in implementation very different from classical product codes. As far as the authors know, classical product codes remained unused in quantum error correction until Hivadi \cite{hivadi2018quantum} showed that if $H$ is the parity check matrix of the 2-fold product of the $[4s,4s-1]$ single parity check code, then there exists a permutation $\pi$ such that $H\pi H^T=0$. This gives a $[[16s^2,16s^2-16s+2,4]]$ quantum CSS code for any $s \in \N$, with the $X$ and $Z$ parity check matrices given by $H$ and $H \pi^\T$ respectively.

\subsection{Classical product construction for CSS codes}

This work introduces a new product code construction which greatly generalizes the one given by Hivadi in~\cite{hivadi2018quantum}. We show that the construction does not apply uniquely to SPC codes, but actually to any family of codes which can be seen as tensor products of other codes. Equipped with this interpretation, for the first time we extend the construction to the $D$-fold product of codes, which allows us to obtain codes of arbitrarily large distances. 

These codes do not technically satisfy the LDPC condition as the stabilizer weight grows unbounded under $D$-fold products for $D\rightarrow \infty$, but in practice the growth is rather slow; on the positive side, the encoding rate converges to 1, as opposed to merely reaching a constant value. Crucially, we show that in the intermediate and practically relevant regime with block sizes involving a few hundred qubits, these codes exhibit superior performance compared to some other codes stemming from true LDPC code families. Furthermore, our codes come with so-called meta-checks naturally embedded in their structure. They are beneficial as they can be used to correct errors that occur in the syndrome measurement readouts, which are always present in practice. This is a step towards single-shot decoding and simplifies the route towards full fault-tolerance. We also show that it is possible to construct codes having different $X$ and $Z$ PCMs, which can yield codes having asymmetric protection against $X$ and $Z$ errors and which might be appealing for certain error models.

Within our general construction, we identify in particular a $[[512,174,8]]$ code that is an especially promising candidate for quantum error correction in the near term. This code combines many desirable properties: high rate; high ability to correct errors; low weight of the syndrome measurements; parallelizable and hardware friendly syndrome measurement circuit; redundancy in the syndrome measurements that guarantees the correction of one syndrome read-out error (even under adversarial noise). We then numerically evaluate the practical performance of this code using the maximum likelihood decoder on the quantum erasure channel and Belief Propagation (BP) without any post-processing on the depolarising channel. This code achieves good performance in both cases, outperforming codes chosen from other well-known families. As far as the authors know, prior work on quantum error correction offers examples that have some of these desirable properties, but no prior example has all of them at the same time.

Finally, we believe that the codes from our product code families may be particularly good candidates to be experimentally implemented, in the near future, in quantum computers based on arrays of Rydberg atoms~\cite{barredo2016atom, gross2017quantum, arute2019quantum, holz2020-2d, wu2021strong}. In fact, there exists a good synergy with this quantum computing platform, as it can realise the transport of an array of atoms and, therefore, can implement a fast fully-parallelised implementation of the syndrome measurements defining our CSS product codes~\cite{xu2024constant, bluvstein2024logical}.

\subsection*{Paper organisation}

Section~\ref{sec:preliminaries} introduces required notation and preliminary information, followed by a description of the new code constructions in \ref{sec:constructions}. A comparison with other codes from the literature is presented in Section~\ref{sec:previous_works}, while the resilience to syndrome errors is discussed in Section~\ref{sec:syndrome_errors}, before concluding the paper in Section~\ref{sec:conclusions}.

\section{Preliminaries and notation}\label{sec:preliminaries}

\subsection{Classical codes}

Let $\F_2$ denote the field with two elements and $\F_2^n$ denote the vector space over $\F_2$ of column vectors with $n$ components, i.e., the set of bit strings of length $n$. Extending the notation further, let $\F_2^{m \times n}$ denote the space of matrices having  $m$ rows and $n$ columns with elements from $\F_2$.

Let $\cC \subseteq \F_2^n$ be a linear vector space which we interpret as the set of all valid code words. We say that $\cC$ is a (binary linear) \textit{code}. If the $\F_2$-linear dimension of $\cC$ is $k$ (there are $2^k$ total code words) we say that $\cC$ encodes $k$ bits in $n$ bits, where $n$ is the \textit{length} of the code, $k$ is the \textit{dimension} of the code and $R = k/n \leq 1$ is the \textit{rate} of the code. The \textit{distance} $d$ of a code $\cC$ is the minimum Hamming distance between any pair of distinct code words and, since $\cC$ is linear, we also have $d = \min_{w \in \cC \setminus \{0\}} |w|$, where $|\cdot|$ denotes the Hamming weight. We summarise the parameters of a classical code $\cC$ with an ordered triple $[n,k,d]$ giving the length, dimension, and minimum distance of the code. Sometimes we omit the code minimum distance and write only $[n,k]$.

A linear code can be defined as the kernel of a {PCM} $H \in \F_2^{m\times n}$, i.e., $ \cC=\{w \in \F_2^n : Hw=0\}$. For a bit string $y\in \F_2^n$ we call $e = Hy$ the \textit{syndrome} associated to $y$. A code can be also specified by a \textit{generator matrix} $G\in \F_2^{n\times k}$ so that the code space is given by the $\F_2$-linear combination of the columns of $G$, that is $\cC = Im(G)$, and the equation $HG = 0$ holds. A PCM $H$ may have \textit{linearly dependent rows}, thus $r = \rank(H) \leq m$ and the code dimension is given as $k = n - r \geq n - m$. Product code constructions typically result in a certain number $\m=m-r$ of linearly dependent (i.e., redundant) checks, which we call \textit{meta-checks}. Product code constructions can result in the introduction of new meta-checks, besides the ones that may be already present in the components codes.

Given $H$ we denote $w_c$ ($w_r$) the maximum Hamming weight of any given column (row) of $H$. We say that a family of PCMs is \textit{sparse} when $w_c$ and $w_r$ are upper bounded by a constant or by a slowly growing function of the code size.

\subsection{Classical product codes ($\cC_\times$)}

\begin{figure}
    \centering
    \hspace{-18mm}
    \scalebox{0.8}{
    \input{01_Fig_ProductCode}
    }
    \caption{Array representation of the $54$ bits of a code given by the product of a $[9, 6]$ and of a $[6, 4]$ systematic code. The final rate is $R = (6/9) (4/6) = 4/9$. The bits on white cells correspond to message bits, bits on hatched cells corresponds to parity-check bits, bits on doubly-hatched cells to checks-of-checks or \textit{meta-checks}.}
    \label{fig:ProductCode}
\end{figure}

Consider two classical codes $\cC_1 \subseteq \F_2^{n_1}$ and $\cC_2 \subseteq \F_2^{n_2}$ having associated PCMs $H_1 \in \F_2^{m_1 \times n_1}$ and $H_2 \in \F_2^{m_2 \times n_2}$. The corresponding $2$-fold product code $\cC_\times$ is defined~\cite{elias1954error} as the code associated to the following PCM $H_\times$,
\begin{align}
\label{eq:H_times}
  H_\times := 
  \begin{pmatrix}
    H_1 \o I \\
	  I \o H_2
  \end{pmatrix}
  \quad \in \
  \F_2^{\,(n_1m_2 + m_1n_2)\times n_1n_2} . 
\end{align}
Above, $I$ denotes identity matrices of appropriate dimensions, i.e., of dimension $n_2$ in the top block and $n_1$ in the bottom block, and $\otimes$ the tensor or Kronecker product. The resulting product code space $\cC_\times$ is given by the tensor product of the component codes, i.e.,
\begin{align}
\label{C_p}
    \cC_\times = \cC_1 \otimes \cC_2  
     \subseteq \
    \F_2^{n_1} \otimes \F_2^{n_2} .
\end{align}
The code dimension is $k_\times = k_1 k_2$, the maximum weight by row and columns can be directly deduced from  \eqref{eq:H_times}, while the code minimum distance is given by $d_\times = d_1 d_2$ (see Appendix~\ref{app:proofs} for a proof of the minimum distance). We refer to~\cite[Chapter 5.7]{peterson1972error} for further details on product codes.

\begin{remark}
A $2$-fold product code is visualized in Figure~\ref{fig:ProductCode}. The  $n_\times = n_1 n_2$ bits are placed in a $2$-dimensional array having $n_1$ columns and $n_2$ rows.
Errors in the product code word can be detected by measuring the syndromes. For each row one can apply the PCM $H_1$ in order to extract the associated syndrome, which is equivalent to applying a PCM given by $H_1 \otimes I$ to all the $n$ bits. Similarly, for each column, one can apply the PCM $H_2$ to extract the associated syndrome. For \textit{systematic} codes (or more precisely systematic encoders), the code words $w \in \F_2^n$ are given by the concatenation of the message $x \in \F_2^k$ and a set of parity-check bits $c \in \F_2^{n-k}$, i.e., $w = (x, c)$. In the corresponding product code, errors are detected by checking whether the last $m_1 = n_1 - k_1$ bits of a row are valid linear combinations of the message bits. The same holds for the last $m_2 = n_2 - k_2$ bits in each column. There are $m_1 m_2$ bits that can be interpreted as checks-of-checks or \textit{meta-checks} (see Figure~\ref{fig:ProductCode}). Meta-checks can be useful in quantum error correcting codes as discussed in Section~\ref{sec:syndrome_errors}.
\end{remark}

The properties of $2$-fold product codes are summarised in the table below, where the subscripts $1$ and $2$ refer to properties of $\cC_1$ and of $\cC_2$, respectively.

\begin{table*}[h]
\centering
\begin{tabular}{|l|c@{\hskip 2pt}l|}
    \hline
    \multicolumn{3}{|c|}{Properties of $\cC_\times$} \\
    \hline \hline
    length        & $n_\times$     &= $n_1 n_2$                 \\ \hline

    checks        & $m_\times$     &= $n_1m_2 + m_1n_2$         \\ \hline

    meta-checks   & $\m_\times$    &= $(m_1-r_1)n_2 + n_1(m_2-r_2) + r_1 r_2$  \\ \hline

    dimension     & $k_\times$     &= $k_1 k_2$                 \\ \hline

    distance      & $d_\times$     &= $d_1 d_2$                 \\ \hline

    row weight    & $w_{\times,r}$ &= $\max(w_{1,r},w_{2,r})$   \\ \hline

    column weight & $w_{\times,c}$ &= $w_{1,c} + w_{2,c}$       \\ \hline

\end{tabular}    
\end{table*}

\subsection{Classical binary tensor product codes ($\cC_\otimes$)}

Let $\cC_1$ and $\cC_2$ be codes having associated PCMs $H_1 \in \F_2^{m_1 \times n_1}$ and $H_2 \in \F_2^{m_2 \times n_2}$. The associated tensor product code $\cC_\otimes$ is defined~\cite{wolf1965codes} as the code originating from the PCM
\begin{align}
    H_\otimes := H_1 \otimes H_2 
    \quad \in \
    \F_2^{m_1 \times n_1} \otimes \F_2^{m_2 \times n_2} 
    \cong 
    \F_2^{m_1m_2 \times n_1n_2}  .
\end{align}
Note that a tensor product code is \textit{not} the tensor product of the component codes: we would obtain this result instead by taking the tensor product of the generator matrices of the codes. Using the fact that a code $\cC$ and its dual $\cC^\perp := \{ w \in \F_2^n \ | \ \langle v , w \rangle = 0, \ \forall v \in \cC \}$ (where $\langle \cdot , \cdot \rangle$ denotes the $\F_2$ inner product) are obtained from each other by interchanging the roles of the parity check and of the generator matrix, we obtain
\begin{align}
    \cC_\otimes := (\cC_1^\perp \otimes \cC_2^\perp)^\perp 
    \quad \subseteq \
    \F_2^{n_1} \otimes \F_2^{n_2} \cong \F_2^{n_1 n_2}.
\end{align}

From the defining properties of the tensor product it follows that $r_\otimes = \rank(H_\otimes)$ is equal to  $r_1r_2 = \rank(H_1) \rank(H_2)$, hence the dimension of the code is $k_\otimes = n_1n_2 - r_1r_2 = k_1n_2+n_1k_2-k_1k_2$. Moreover, one can immediately verify that the row and column weights of $H_\otimes$ are simply given by the product of the individual row and column weights of the component codes. Tensor product codes do not have meta-checks, unless some are already present in the component codes. Finally, the distance of the code is $d_\otimes = \min (d_1,d_2)$ which is discussed in Appendix~\ref{app:proofs} for completeness. See~\cite{wolf2006introduction} for a short introduction to the topic and the extension to non-binary component codes. 

In summary, tensor product constructions allow to create codes having higher rates than the component codes, at the cost of increasing the PCM weights and no improvement in the code distance. The properties of the tensor product code are summarised in the table below.

\begin{center}
\begin{tabular}{|l|c@{\hskip 2pt}l|}
    \hline
    \multicolumn{3}{|c|}{Properties of $\cC_\otimes$}     \\
    \hline \hline
    length        & $n_\otimes$ & $= n_1 n_2$             \\ \hline

    checks        & $m_\otimes$ & $= m_1 m_2$             \\ \hline

    meta-checks   & $\m_\otimes$ & $= m_1 m_2 - r_1 r_2$  \\ \hline

    dimension     & $k_\otimes$ & $= n_1n_2-r_1r_2$       \\ \hline

    distance      & $d_\otimes$ & $= \min(d_1, d_2)$      \\ \hline

    row weight    & $w_{\otimes,r}$ & $= w_{1,r} w_{2,r}$ \\ \hline

    column weight & $w_{\otimes,c}$ & $= w_{1,c} w_{2,c}$ \\ \hline

\end{tabular}    
\end{center}

\subsection{Quantum CSS codes}

Let $\C(\F_2^n) \cong \C^{2^n}$ be the complex Hilbert space associated to the space of $n$ qubits, equipped with an orthonormal basis. Each of the $2^n$ basis vectors is indexed by an element of $\F_2^n$. The Pauli operators are $\cP = \{I, X, Y, Z\}$ (with $Y = iXZ$) and the $n$-qubit Pauli group $\cP_n$ acting on $\C(\F_2^n)$ is obtained by taking tensor products of $n$ Pauli operators, together with a global phase $\alpha \in \{\pm 1, \pm i\}$. That is, an element of $\cP_n$ has the form $P = \alpha P_1 \otimes \cdots \otimes P_n$, where $P_j \in \{I,X,Y,Z\}$ for all $j$. stabilizer codes are defined by a commutative subgroup $\cS \subseteq \cP_n$ such that $-I \notin \cS$. The quantum code space is a $\C$-linear subspace $\cC \subseteq \C(\F_2^n)$ defined as the set of quantum states that are stabilised by all the elements of $\cS$, that is $\cC := \{\ket{\psi} \in  \C(\F_2^n) \ | \ s\ket{\psi} = \ket{\psi}, \ \forall s \in \cS \}$. The stabilizer group has $2^r$ elements, can be generated by $r$ independent elements, and defines a quantum code $\cC$ having $\C$-dimension $2^{n-r} = 2^k$. In analogy with the classical case, we say that $k$ is the code dimension, $n$ the code length, and $R = k/n$ the code rate.

A CSS code is a stabilizer code where the stabilizer generators can be chosen to be either tensor products of $I$ and $X$ operators only, or tensor products of $I$ and $Z$ operators only. One can thus represent a CSS code with two binary PCMs: $H^z \in \F_2^{m^z \times n}$, which is associated to parity checks in the $Z$ (computational) basis, and $H^x \in \F_2^{m^x \times n}$, which is associated to parity checks in the $X$ (Hadamard) basis. For instance, the stabilizer elements $X \otimes X \otimes I$ corresponds to the row vector $(1,1,0)$ in $H^x$ and $Z \otimes I \otimes Z$ to the row vector $(1,0,1)$ in $H^z$. 
The dimension of the resulting binary code is given by $k = n-r^x - r^z = k^x + k^z - n$, where $r^x = \rank(H^x) \leq m^x$, $r^z = \rank(H^z) \leq m^z$, $k^x = \dim(\ker(H^x))$ and $k^z = \dim(\ker(H^z))$. Commutativity of the generators is equivalent to requiring
\begin{align}
\label{eq:CSS}
    H^x (H^z)^\T = 0   \mod 2.
\end{align}
Neglecting the global phase, an error pattern $P = P_1 \otimes \cdots \otimes P_n \in \cP_n$ is in one to one correspondence with an ordered pair of vectors $(v^z,v^x)$, with $v^z, v^x \in \F_2^n$: we have $[v^z]_j = 1$ if $P_j \in \{Y,Z\}$ and $[v^z]_j = 0$ otherwise; similarly, we have $[v^x]_j = 1$ if $P_j \in \{X,Y\}$ and $[v^x]_j = 0$ otherwise. Note that $X$ checks (specified via $H^x$) are sensitive to $Z$ errors and that $Z$ checks (specified via $H^z$) are sensitive to $X$ errors.

For quantum CSS codes, one can distinguish different kinds of code minimum distance which we refer to as the \textit{pure distance} $\delta$ and the \textit{code distance} $d$. The pure distances for $X$ and $Z$ errors are given, respectively, by 
\begin{align}
\label{pure_dist_x}
    \delta^x & := \min \big\{ |v^x| \ \big| \ H^z v^x = 0, v^x \neq 0 \big\} \\
\label{pure_dist_z}
    \delta^z & := \min \big\{ |v^z| \ \big| \ H^x v^z = 0, v^z \neq 0 \big\}
\end{align}
and we simply call $\delta := \min(\delta^x,\delta^z)$ the pure distance. By contrast, the code distances for $X$ and $Z$ errors are obtained by considering only errors patterns that do not leave the code-subspace invariant and thus do result in a logical error:
\begin{align}
\label{dist_x}
    d^x & := \min \big\{ |v^x| \ \big| \ 
        H^z v^x = 0, v^x \notin \vspan \!\big( (H^x)^\T \big) \big\} \\
\label{dist_z}
    d^z & := \min \big\{ |v^z| \ \big| \ 
        H^x v^z = 0, v^z \notin \vspan \!\big( (H^z)^\T \big) \big\}
\end{align}
We call $d:= \min(d^x,d^z)$ the code distance. Note that we always have $d \geq \delta$ and $d$ can be significantly larger than $\delta$ in certain cases\footnote{For instance, the surface code has constant pure distance, $\delta = 4$, while the code distance scales as the square root of the length $d = O(\sqrt{n})$.}. We summarise the parameters of a quantum code $\cC$ with the notation $[[n,k,d]]$ or, omitting the code distance, $[[n,k]]$.

\section{Main constructions}\label{sec:constructions}

We first present a method for obtaining asymmetric and symmetric 2-fold products of CSS codes. Then, we generalise the symmetric construction to $D$-fold products. Classical product and tensor product constructions are employed in such a way that the commutativity condition for CSS codes is automatically fulfilled.

\subsection{Asymmetric $2$-fold product construction of CSS codes ($\cC_\ltimes$)}

\begin{defin}
Let $\cC_1$ and $\cC_2$ be CSS codes having, for $\ell \in \{1,2\}$, PCMs associated to $X$ and $Z$ checks given by $H_{\ell}^x \in \F_2^{m_\ell^x\times n_\ell}$ and by $H_{\ell}^z \in \F_2^{m_\ell^z\times n_\ell}$. Because of the commutativity condition, these PCMs satisfy:
\begin{align}
\label{CSS_ap}
    H_{\ell}^x (H_{\ell}^z)^\T = 0 
	\qquad \ell \in \{1,2\}
\end{align}

We define the asymmetric 2-fold product quantum CSS code $\cC_\ltimes$ as the code associated to the following $X$ and $Z$ parity check matrices:
\begin{align}
  H^x_\ltimes
  & := 
    \begin{pmatrix}
	  H^x_1 \o I \\
	  I \o H^x_2        
    \end{pmatrix}
  \\
  H^z_\ltimes
  & := 
  \Big(\!\!
    \begin{array}{ccc}
	  H^z_1 \o H^z_2
    \end{array}
  \!\!\Big) .
\end{align}
\end{defin}

We have given the definition of $\cC_\ltimes$ for two arbitrary CSS codes for sake of generality. We note that asymmetric product codes as here defined typically provide more protection against $Z$ errors than $X$ errors. This may be beneficial for error channels where $Z$ errors occur more frequently than $X$ errors~\cite{aliferis2008fault, martinis2009superconducting, guillaud2019repetition, chamberland2022building}. Alternatively, the construction methods for the $X$ and $Z$ PCMs may be reversed. This CSS code features meta-checks in the PCM based on the product code construction but no meta-checks in PCM based on the tensor-product code construction. Finally, the commutativity condition $H^x_\ltimes (H^z_\ltimes)^\T = 0$ can be verified using Eq.~\eqref{CSS_ap} and the multi-linearity of tensor products (which implies $0 \otimes M = 0$ for any matrix $M$). 

Expressions for the pure distances can be easily derived from the distances of the underlying codes and the properties of  the classical product and tensor product construction. The true code distances [see the definitions given in Eq.~\eqref{dist_x} and \eqref{dist_z}] cannot be easily computed, but can only be lower bounded using $d^x \geq \delta^x$ and $d^z \geq \delta^z$. In contrast to the classical case, the distance of $\cC_\ltimes$ can be smaller than the product of the distances of the component codes; for instance, $d_\ltimes^z < d_1^z d_2^z$, as we show in Appendix~\ref{app:proofs}.

In the table below, we report the properties of $\cC_\ltimes$ both for the general case (left column) and for the special case in which the two component codes are equal (right column). These properties can be straightforwardly derived from the classical product and tensor product constructions. The dimension of the CSS code $\cC_\ltimes$ is most easily expressed in terms of the quantities $k_\ell^x = \dim(\ker(H_\ell^x)), r_\ell^z = \rank(H_\ell^z)$.

\begin{center}
\begin{tabular}{|l|c@{\hskip 2pt}l|c@{\hskip 2pt}l|}
    \hline
    \multicolumn{5}{|c|}{Properties of $\cC_\ltimes$} \\
    \hline \hline
    
    length
    & $n_\ltimes$ & $= n_1 n_2$  
    & $n_\ltimes$ & $= n^2$ 
    \phantom{\Big\vert} \\ \hline
    
    $X$ checks
    & $m_\ltimes^x$ & $= m_1^x n_2 + n_1 m_2^x$  
    & $m_\ltimes^x$ & $= 2n m^x$ \\
    
    $Z$ checks
    & $m_\ltimes^z$ & $= m_1^z m_2^z$  
    & $m_\ltimes^z$ & $= (m^z)^2$\\ \hline
    
    $X$ meta-checks
    & $\m_\ltimes^x$ & $=(m_1^x-r_1^x)n_2 + n_1 (m_2^x-r_2^x) + r_1^xr_2^x $
    & $\m_\ltimes^x$ & $=2n(m^x-r^x) + (r^x)^2 $ 
    \phantom{\Big\vert} \\ \hline

    dimension
    & $k_\ltimes$ & $= k_1^x k_2^x - r_1^z r_2^z$
    & $k_\ltimes$ & $= (k^x)^2 - (r^z)^2$
    \phantom{\Big\vert} \\ \hline
    
    $Z$ pure distance
    & $\delta_\ltimes^z$ & $ = \delta^z_1 \delta^z_2 $     
    & $\delta_\ltimes^z$ & $ = (\delta^z)^2$  \\
    
    $X$ pure distance
    & $\delta_\ltimes^x$ & $ = \min(\delta^x_1,\delta^x_2)$        
    & $\delta_\ltimes^x$ & $ = \delta^x$ \\ \hline
    
    $X$ row weight
    & $w_{\ltimes,r}^x$ & $ = \max(w_{1,r}^x, w_{2,r}^x)$
    & $w_{\ltimes,r}^x$ & $ = w_{r}^x$ \\ 
    
    $Z$ row weight
    & $w_{\ltimes,r}^z$ & $ = w_{1,r}^z w_{2,r}^z$
    & $w_{\ltimes,r}^z$ & $ = (w_{r}^z)^2 $  \\ \hline
    
    $X$ column weight
    & $w_{\ltimes,r}^x$ & $ = w_{1,c}^x + w_{2,c}^x$
    & $w_{\ltimes,r}^x$ & $ = 2w_{c}^x$ \\ 
    
    $Z$ column weight
    & $w_{\ltimes,c}^z$ & $ = w_{1,c}^z w_{2,c}^z$
    & $w_{\ltimes,c}^z$ & $ = (w_{c}^z)^2 $  \\ \hline
\end{tabular}    
\end{center}

\subsection{Symmetric 2-fold product construction of CSS codes ($\cC_\boxtimes$)}

\begin{defin}

Let $\cC_1,\cC_2,\cC_3,\cC_4$ be a set of four CSS codes such that, for each $\ell \in \{1,2,3,4\}$, the PCMs associated to $X$ and $Z$ checks are $H_{\ell}^x \in \F_2^{m_\ell^x\times n_\ell}$ and  $H_{\ell}^z \in \F_2^{m_\ell^z\times n_\ell}$. Because of the commutativity condition, these PCMs must satisfy:
\begin{align}
\label{CSS_qp}
	H_{\ell}^x (H_{\ell}^z)^\T = 0 
	\qquad  \ell \in \{1,2,3,4\}
\end{align}

We define the symmetric 2-fold product quantum CSS code $\cC_\boxtimes$ as the code associated to the following $X$ and $Z$ parity check matrices:
\begin{align}
\label{PCM_2d_X}
    H^x_\boxtimes & := 
    \begin{pmatrix}
        H_{1}^x \o H_{2}^x \o I \o I \\
        I \o I \o H_{3}^x \o H_{4}^x
    \end{pmatrix}
    \\
\label{PCM_2d_Z}
    H^z_\boxtimes & := 
    \begin{pmatrix}
    	  H_{1}^z \o I \o H_{3}^z \o I \\
    	  I \o H_{2}^z \o I \o H_{4}^z
    \end{pmatrix} .
\end{align}
\end{defin}

Note that both the $X$ parity checks and $Z$ checks are given by classical product codes. Specifically, these are the products of component codes associated to $H_{1}^x \otimes H_{2}^x$ and $H_{3}^x \otimes H_{4}^x$ (for $H^x_\boxtimes$) and $H_{1}^z \otimes H_{3}^z$ and $H_{2}^z \otimes H_{4}^z$ (for $H^z_\boxtimes$). This implies, in particular, that $H^x_\boxtimes$ and $H^z_\boxtimes$ are not full-rank. One can immediately verify the commutativity condition $H^x_\boxtimes (H^z_\boxtimes)^\T = 0$ using Eq.~\eqref{CSS_qp}. 

The properties of the quantum CSS code $\cC_\boxtimes$ are given in the table below, both for the general case and for the special case where the component codes are equal; the dimension of the code is computed via the equation $k_\boxtimes = n_\boxtimes - \rank(H_\boxtimes^x) - \rank(H_\boxtimes^z)$ and expressed in terms of the quantities $r_\ell^x = \rank(H_\ell^x), r_\ell^z = \rank(H_\ell^z)$.
\begin{small}
\begin{center}
\begin{tabular}{|l|c@{\hskip 2pt}l|c@{\hskip 2pt}l|}
    \hline
    \multicolumn{5}{|c|}{Properties of $\cC_\boxtimes$} \\
    \hline \hline
    
    length
    & $n_\boxtimes$ & $= n_1 n_2 n_3 n_4$  
    & $n_\boxtimes$ & $= n^4$ 
    \phantom{\Big\vert} \\ \hline
    
    $X$ checks
    & $m_\boxtimes^x$ & $= m_1^x m_2^x n_3n_4 + n_1n_2 m_3^x m_4^x$  
    & $m_\boxtimes^x$ & $= 2n^2 (m^x)^2$ \\
    
    $Z$ checks
    & $m_\boxtimes^z$ & $= m_1^z n_2 m_3^z n_4 + n_1 m_2^z n_3 m_4^z$  
    & $m_\boxtimes^z$ & $= 2n^2 (m^z)^2$\\ \hline
    
    $X$ meta-checks
    & $\m_\boxtimes^x$ & $=(m_1^x m_2^x - r_1^x r_2^x) n_3 n_4 + n_1 n_2 (m_3^x m_4^x - r_3^x r_4^x)$ & $\m_\boxtimes^x$ & $=2 n^2 ((m^x)^2-(r^x)^2)$  \\ 
    & & $+\; r_1^x r_2^x r_3^x r_4^x$ & & $+\; (r^x)^4$      \\
    
    $Z$ meta-checks
    & $\m_\boxtimes^z$ & $=(m_1^z m_3^z - r_1^z r_3^z) n_2 n_4 + n_1 n_3 (m_2^z m_4^z - r_2^z r_4^z)$ & $\m_\boxtimes^z$ & $=2 n^2 ((m^z)^2-(r^z)^2)$  \\ 
    & & $+\; r_1^z r_2^z r_3^z r_4^z$ & & $+\; (r^z)^4$      \\
    \hline
    
    dimension
    & $k_\boxtimes$ & $= \sprod_{\ell} n_\ell + \sprod_\ell r^x_\ell + \sprod_\ell r^z_\ell$ & $k_\boxtimes$ & $= n^4 + (r^x)^4 + (r^z)^4$ \\ & & $-\, r_1^x r_2^x n_3 n_4 - n_1 n_2 r_3^x r_4^x - r_1^z n_2 r_3^z n_4 - n_1 r_2^z n_3 r_4^z$ & & $-\, 2 n^2 (r^x)^2 - 2 n^2 (r^z)^2$  
    \phantom{\Big\vert} \\ \hline
    
    $Z$ pure distance
    & $\delta_\boxtimes^z$ & $ = \min(\delta^z_1,\delta^z_2) \, \min(\delta^z_3,\delta^z_4)$     
    & $\delta_\boxtimes^z$ & $ = (\delta^z)^2$  \\
    
    $X$ pure distance
    & $\delta_\boxtimes^x$ & $ = \min(\delta^x_1,\delta^x_3) \, \min(\delta^x_2,\delta^x_4)$        
    & $\delta_\boxtimes^x$ & $ = (\delta^x)^2$ \\ \hline
    
    $X$ row weight
    & $w_{\boxtimes,r}^x$ & $ = \max(w_{1,r}^x w_{2,r}^x, w_{3,r}^x w_{4,r}^x)$
    & $w_{\boxtimes,r}^x$ & $ = (w_{r}^x)^2 $ \\ 
    
    $Z$ row weight
    & $w_{\boxtimes,r}^z$ & $ = \max(w_{1,r}^z w_{3,r}^z, w_{2,r}^z w_{4,r}^z)$
    & $w_{\boxtimes,r}^z$ & $ = (w_{r}^z)^2 $  \\ \hline
    
    $X$ column weight
    & $w_{\boxtimes,r}^x$ & $ = w_{1,c}^x w_{2,c}^x + w_{3,c}^x w_{4,c}^x$
    & $w_{\boxtimes,r}^x$ & $ = 2(w_{c}^x)^2 $ \\ 
    
    $Z$ column weight
    & $w_{\boxtimes,c}^z$ & $ = w_{1,c}^z w_{3,c}^z + w_{2,c}^z w_{4,c}^z$
    & $w_{\boxtimes,c}^z$ & $ = 2(w_{c}^z)^2 $  \\ \hline
\end{tabular}    
\end{center}
\end{small}
\begin{remark}
We note that there is some arbitrariness in our choice of how to present the PCMs of the product code. For instance, we may choose as $Z$ PCM
\begin{align}
    H'^z_\boxtimes := 
    \begin{pmatrix}
    	  H_{1}^z \o P_2 \o P_3 \o H_{4}^z \\
    	  P_1 \o H_{2}^z \o H_{3}^z \o P_4        
    \end{pmatrix}
\end{align}
where $P_1,P_2,P_3,P_4$ are permutation matrices of appropriate size,
while keeping the $X$ PCM unchanged. However, this definition is equivalent to the one given in Eq.~\eqref{PCM_2d_X}, up to a re-labelling of the qubits and of the stabilizer generators.
\end{remark}

\subsection{Symmetric $\textbf{D}$-fold product construction of CSS codes ($\cC_\textbf{(D)}$)}

The construction of symmetric product CSS codes can be easily generalised to higher dimensions, by taking the product of $D^2$ CSS codes. For instance, by using the PCMs of a set of $9$ CSS codes one can construct the PCMs of a $3$-fold product code as follows:
\begin{align}
\label{PCM_3d_x}
  & H_{(3)}^x := 
  \left(
    \begin{array}{ccccccccccccccccc}
	  H_{1}^x \o H_{2}^x \o H_{3}^x \o I \o I \o I \o I \o I \o I\\
	  I \o I \o I \o H_{4}^x \o H_{5}^x \o H_{6}^x \o I \o I \o I\\ 
	  I \o I \o I \o I \o I \o I \o H_{7}^x \o H_{8}^x \o H_{9}^x
    \end{array}
  \right)
  \\
\label{PCM_3d_z}
  & H_{(3)}^z := 
  \left(
    \begin{array}{ccccccccccccccccc}
	  H_{1}^z \o I \o I \o H_{4}^z \o I \o I \o H_{7}^z \o I \o I\\
	  I \o H_{2}^z \o I \o I \o H_{5}^z \o I \o I \o H_{8}^z \o I\\ 
	  I \o I \o H_{3}^z \o I \o I \o H_{6}^z \o I \o I \o H_{9}^z
    \end{array}
  \right) .
\end{align}
It is easy to check that it is a CSS code and its properties can be derived similarly as done for the case of a 2-fold product. The general definition of the symmetric D-fold product construction of CSS codes is as follows.

\begin{defin} 
\label{def:D-fold}
Let $\cC_1, \ldots ,\cC_{D^2}$ be a set of $D^2$ CSS codes and, for each $\ell \in \{1,\ldots,D^2\}$, let $H_{\ell}^x \in \F_2^{m_\ell^x\times n_\ell}$ and $H_{\ell}^z \in \F_2^{m_\ell^z\times n_\ell}$ be the PCMs associated to $X$ and $Z$ checks, respectively. We define the symmetric $D$-fold product CSS code $\cC_{(D)}$ as the code associated to the following $X$ and $Z$ parity check matrices:
\begin{align}
\label{PCM_D_X}
    H^x_{(D)} & := 
    \overset{D-1}{\underset{j=0} {{\Big[}\textup{Stack}{\Big]}} }
    \bigotimes_{\ell=1}^{D^2} \mathcal{H}_{\ell,j}^x
    \qquad
    \mathcal{H}_{\ell,j}^x = 
    \begin{cases} 
    H_\ell^x   & \text{if } jD+1 \leq \ell \leq (j+1)D \\
    I_{n_\ell} & \text{otherwise}
    \end{cases}
    \\
\label{PCM_D_Z}
    H^z_{(D)} & := 
    \overset{D-1}{\underset{j=0} {{\Big[}\textup{Stack}{\Big]}} }
    \bigotimes_{\ell=1}^{D^2} \mathcal{H}_{\ell,j}^z
    \qquad
    \mathcal{H}_{\ell,j}^z = 
    \begin{cases} 
    H_\ell^z   & \text{if } (\ell-1) \equiv j \mod D\\
    I_{n_\ell} & \text{otherwise}
    \end{cases}
\end{align}
where $\textup{Stack}$ denotes the operation of creating an $m\times n$ matrix ($m = \sum_j m_j$) by stacking $m_j \times n$ matrices one above the other.
\end{defin}

Other equivalent definitions of the code $\cC_{(D)}$ could be given by choosing different orderings of the matrices $H_\ell^x$ and $H_\ell^z$ in the tensor products. The properties of this code can be obtained with similar procedures as in the previous examples. For instance, we can compute the number of meta-checks $\m_{(D)}^\alpha$, for $\alpha \in \{x,z\}$, by 
\begin{align}
    \m_{(D)}^\alpha = m_{(D)}^\alpha - \left(n_{(D)} - k_{(D)}^\alpha\right) 
    \qquad \alpha \in \{x,z\}\,,
\end{align}
where $k_{(D)}^x$ and $k_{(D)}^z$ are the classical code dimensions associated to the $X$ and $Z$ PCMs. These can be computed as a function of $n_\ell$ and $r_\ell^\alpha$ using the definitions of $H_{(D)}^\alpha$ in \eqref{PCM_D_X} and \eqref{PCM_D_Z}, resulting in
\begin{align}
    k_{(D)}^x
    & =
    \prod_{j=0}^{D-1}
    \Bigg(
    \prod_{\ell=jD+1}^{(j+1)D} n_\ell - 
    \prod_{\ell=jD+1}^{(j+1)D} r_\ell^x
    \Bigg) \\
    k_{(D)}^z
    & =
    \prod_{j=0}^{D-1}
    \Bigg(
    \prod_{\substack{(\ell-1) \equiv j \\ \mod D~~}} n_\ell -
    \prod_{\substack{(\ell-1) \equiv j \\ \mod D~~}} r_\ell^z
    \Bigg) .
\end{align}

We summarise the properties of this family of codes for the special case where all the component codes are equal (the expressions for the general case are lengthy and can be derived straightforwardly).

\begin{center}
\begin{tabular}{|l|c@{\hskip 2pt}l|}
    \hline
    \multicolumn{3}{|c|}{Properties of $\cC_{\textup{(D)}}$} \\
    \hline \hline
    
    length
    & $n_{(D)}$ & $= n^{D^2} $ 
    \phantom{\Big\vert} \\ \hline
    
    $X$ checks
    & $m_{(D)}^x$ & $= D\, n^{D(D-1)}\, (m^x)^D$ \\
    
    $Z$ checks
    & $m_{(D)}^z$ & $= D\, n^{D(D-1)}\, (m^z)^D$ \\ \hline
    
    $X$ meta-checks
    & $\m_{(D)}^x$ & $= D\, n^{D(D-1)}\, (m^x)^D + (n^D - (r^x)^D)^D - n^{D^2}$ \\
    
    $Z$ meta-checks
    & $\m_{(D)}^z$ & $= D\, n^{D(D-1)}\, (m^z)^D + (n^D - (r^z)^D)^D - n^{D^2}$ \\ \hline
    
    dimension
    & $k_{(D)}$ & $= (n^D- (r^x)^D)^D + (n^D- (r^z)^D)^D - n^{D^2}$
    \phantom{\Big\vert} \\ \hline
    
    $Z$ pure distance
    & $\delta_{(D)}^z$ & $= (\delta^z)^D$ \\
    
    $X$ pure distance
    & $\delta_{(D)}^x$ & $= (\delta^x)^D$ \\ \hline
    
    $X$ row weight
    & $w_{(D),r}^x$ & $ = (w_{r}^x)^D$ \\
    
    $Z$ row weight
    & $w_{(D),r}^z$ & $ =(w_{r}^z)^D$ \\ \hline
    
    $X$ column weight
    & $w_{(D),c}^x$ & $ = D (w_{c}^x)^D$ \\
    
    $Z$ column weight
    & $w_{(D),c}^z$ & $ = D (w_{c}^z)^D$ \\ \hline
\end{tabular}    
\end{center}

\subsection{Single-parity-check $\textbf{D}$-fold product CSS codes ($\cC_{\textbf{SPC(D)}}$)}
\label{sec:d-fold}

We now specialise the symmetric D-fold product construction to the case where the component codes are all given by single-parity-check (SPC) codes. The choice of SPC codes as component CSS codes is motivated by the fact that the length of $D$-fold product codes grows very fast in $D$ (as $n^{D^2}$), thus reasonably sized product codes can be obtained only for very small sizes of the component codes. A favourable choice is then $H_{\ell}^x = H_{\ell}^z = \begin{pmatrix} 1 &\!\! 1 \end{pmatrix}\ \forall\ell$, 
corresponding to the 2-qubit code having stabilizer generators $X \otimes X$ and $Z \otimes Z$. Note that this CSS code has zero encoding rate, since the only quantum state in its code space is the Bell state $\ket{\Phi^+} =\frac{1}{\sqrt{2}} (\ket{0,0} + \ket{1,1})$. Nonetheless, the $D$-fold product CSS codes obtained from it have a positive encoding rate for all $D \geq 2$. We thus arrive at the following definition.

\begin{defin}
Consider the following set of PCMs for the component codes $\cC_1, \ldots, \cC_{D^2}$:
\begin{align}
\label{eq:SPC(D)}
    H^x_{\ell} = H^z_{\ell}
    := 
    \begin{cases} 
    (\underbrace{1 ~ \cdots ~ 1}_{2s})  
        & \text{for } \ell=(i-1)D+i, \; i=1, \dots, D\\
    (\,1 ~~ 1\,)
        & \text{otherwise}
    \end{cases}
\end{align}
and employ them to construct a $D$-fold product CSS code as described in Definition~\ref{def:D-fold}. We call the resulting code a single-parity-check $D$-fold product CSS code and denote it either $\cC_{\mathrm{SPC}(D,s)}$ or more simply $\cC_{\mathrm{SPC}(D)}$ in the case $s=1$.  
\end{defin}

We remark here that in the special case $D=2$, $\cC_{\mathrm{SPC}(2,s)}$ is equivalent up to qubit permutations to the $[[16s^2,16s^2-16s+2,4]]$ quantum CSS code introduced in~\cite[Theorem 2]{hivadi2018quantum}. 

Note that the codes associated to $X$ and $Z$ checks are the same up to permutations of the columns of $H^x$ and $H^z$, and are isomorphic to the $D$-fold product of the $[s2^D,s2^D-1]$ SPC code. The free parameter $s\in \N$ may be used to increase the rate. The SPC$(D,s)$ code family features a pure distance that grows asymptotically with $D$ and, moreover, the distance is equal to the pure distance, $d = \delta = 2^D$, as we prove in Appendix~\ref{app:proofs}.

The properties of this code family are summarised in the table below. 

\begin{center}
\begin{tabular}{|l|r@{\hskip 2pt}l|}
    \hline
    \multicolumn{3}{|c|}{Properties of $\cC_{\mathrm{SPC}(D,s)}$} \\
    \hline \hline
    
    length
    & $n~$ & $= (s2^D)^D$ 
    \phantom{\Big\vert} \\ \hline
    
    $X$ and $Z$ checks
    & $m^x = m^z$ & $= D (s2^D)^{D-1}$ 
    \phantom{\Big\vert} \\ \hline
    
    $X$ and $Z$ meta-checks
    & $\m^x = \m^z$ & $= D (s2^D)^{D-1} + (s2^D-1)^D-(s2^D)^D$ 
    \phantom{\Big\vert} \\ \hline
    
    dimension
    & $k~$ & $= 2(s2^D-1)^D-(s2^D)^D$
    \phantom{\Big\vert} \\ \hline
    
    $X$ and $Z$ distance
    & $\delta^x = \delta^z$ & $= d^x = d^z = 2^D$ 
    \phantom{\Big\vert} \\ \hline
    
    $X$ and $Z$ row weight
    & $w_{r}^x = w_{r}^z$ & $ = s2^D$ 
    \phantom{\Big\vert} \\ \hline    
    
    $X$ and $Z$ column weight
    & $w_{c}^x = w_{c}^z$ & $ = D$
    \phantom{\Big\vert} \\ \hline
\end{tabular}    
\end{center}

\section{Comparison with previous works}
\label{sec:previous_works}

In this section, we compare the performance of our codes against the performance of other code families. 

We start by selecting the SPC$(3)$ code, i.e., the CSS code from the family described in Section~\ref{sec:d-fold} with parameters $D=3, s=1$. More explicitly, this is the CSS code associated to the following 3-fold product codes:
\begin{align}
\label{3D-SPC_x}
  & H_{\mathrm{SPC}(3)}^x = 
  \left(
    \begin{array}{ccccccccccccccccc}
	  h \o h \o h \o I \o I \o I \o I \o I \o I\\
	  I \o I \o I \o h \o h \o h \o I \o I \o I\\ 
	  I \o I \o I \o I \o I \o I \o h \o h \o h
    \end{array}
  \right)
  \\
\label{3D-SPC_z}
  & H_{\mathrm{SPC}(3)}^z = 
  \left(
    \begin{array}{ccccccccccccccccc}
	  h \o I \o I \o h \o I \o I \o h \o I \o I\\
	  I \o h \o I \o I \o h \o I \o I \o h \o I\\ 
	  I \o I \o h \o I \o I \o h \o I \o I \o h
    \end{array}
  \right)
\end{align}
where $h=\begin{pmatrix} 1 &\!\! 1 \end{pmatrix}$ and $I = \left(\begin{smallmatrix} 1 & 0 \\ 0 & 1 \end{smallmatrix}\right)$. This results in a $[[512,174,8]]$ CSS code where all row weights are 8 and all column weights are 3; each matrix has 23 linearly dependent rows (i.e., meta-checks).

\subsection{Previous code families participating in the comparison}

Next, we search for codes from other families having similar parameters to the SPC$(3)$ code. 

\paragraph{Bicycle codes~\cite{mackay2004sparse}\hspace{-1mm}} were one of the first families of quantum LDPC codes to be proposed. 
A bicycle code is obtained by first randomly selecting a $n/2 \times n/2$ cyclic matrix $C$, which in our case has row weight 8.
Next, the matrix $H_0 =  \begin{pmatrix} C & C^{\T} \end{pmatrix}$ is constructed.
Then, we proceed to remove $k/2$ rows from $H_0$ in order to obtain a matrix $H_1$ with $(n-k)/2$ rows. Finally, the PCM of the bicycle code is obtained by setting $H^x=H^z=H_1$.
Different strategies can be followed in order to remove rows from $H_0$ to obtain $H_1$. In our case, we follow an approach that aims at obtaining a matrix $H_1$ whose column weights are as uniform as possible. In particular, we follow a greedy approach in which rows are removed one by one, and the row to be removed is always selected so as to keep the column weights as uniform as possible. A bicycle code with parameters [[512,174,2]] is then obtained. The block length and rate are chosen to be equal to those of the SPC(3) code, but for this high encoding rate we could not find codes having minimum distance larger than 2.

\paragraph{Hypergraph product codes~\cite{tillich2013quantum}} were introduced as a generalization of the surface code; thus the surface code can be viewed as a hypergraph product of a certain representation of the repetition code with itself. More generally,~\cite[Theorem 1]{tillich2013quantum} gives the following construction: if $H$ is a full rank $m \times n$ PCM of a $[n,k,d]$ classical linear code, then the quantum CSS code with parity check matrices $H^x = \begin{pmatrix} H \otimes I & I \otimes H^{\T} \end{pmatrix}$ and $H^z = \begin{pmatrix} I \otimes H & H^{\T} \otimes I \end{pmatrix}$ has parameters $[[n^2 + m^2, k^2, d]]$. For the present comparison, we want a hypergraph product code with parameters as close as possible to $[[512,174]]$; the closest that one can get within this family is for $n=21, m=8, k=13$, giving a $[[505,169]]$ quantum CSS code. It is known that the distance of a $[21,13]$ classical linear code is at most 4 \cite{Grassl:codetables}. Using a computer search, we found a sparse $[21,13,4]$ classical code that leads to a $[[505,169,4]]$ hypergraph product code. The rows of the parity check matrices have weights between 9 and 13. Most of the columns have weight 3, but some have higher weights up to 10. 

\paragraph{Quantum Tanner codes~\cite{leverrier2022quantum}} were introduced as an asymptotically good family of quantum LDPC codes. Recently, \cite{gu2023efficient,leverrier2022efficient,dinur2023goodquantum,leverrier2023decoding} have discussed efficient decoders for these codes. There are slight variations in the construction of quantum Tanner codes between the various references. For the present comparison, we adopt the approach of~\cite{leverrier2022efficient,leverrier2023decoding} which avoids the total non-conjugacy constraint.

A quantum Tanner code is described by seven components: a finite group $G$, two subsets $A,B \subset G$ of size $\Delta$ each that are closed under inversion, two full rank $m \times \Delta$ parity check matrices $H_A, H_B^\bot$ and two full rank $(\Delta - m) \times m$ parity check matrices $H_A^\bot, H_B$, where $H_A, H_A^\bot$ are orthogonal, and so are $H_B, H_B^\bot$, and where $m \leq \Delta/2$ is a parameter that determines a lower bound $(1-2m/\Delta)^2$ on the rate. 

The group $G$ and its two subsets $A,B$ define an incidence structure with vertices, edges and squares called a left-right Cayley complex~\cite{dinur2022locally}. In the approach of~\cite{leverrier2022efficient,leverrier2023decoding}, the vertices are 
\begin{align}
    V= \big\{ (g,i,j) \;\big\vert\; g \in G, i \in \{0,1\}, j \in \{0,1\} \big\}
\end{align}
the edges are $E =  E_A \cup E_B$ with
\begin{align}
    E_A & = \big\{ ((g,i,0),(ag,i,1)) \;\big\vert\; g \in G, a \in A, i \in \{0,1\} \big\} \\ 
    E_B & = \big\{ ((g,0,j),(gb,1,j)) \;\big\vert\; g \in G, b \in B, j \in \{0,1\} \big\}
\end{align}  and the squares are 
\begin{align}
    Q = \big\{ ((g,0,0),(ag,0,1),(agb,1,1),(gb,1,0)) \;\big\vert\; g \in G, a \in A, b \in B \big\} .
\end{align}
Qubits are placed on the squares and constraints are placed on the vertices. If $i+j \mod 2 =0$, vertex $(g,i,j)$ imposes the $x$ constraints with PCM $H_A \otimes H_B$ on the connected squares/qubits, and if $i+j \mod 2 = 1$, then vertex $(g,i,j)$ imposes the $z$ constraints with PCM $H_A^\bot \otimes H_B^\bot$ on the connected squares/qubits. 

The resulting quantum CSS code has $\Delta^2 |G|$ qubits and $2m(\Delta-m)|G|$ rows in each of its parity check matrices $H^x,H^z$. The quality of the code is determined by the spectral expansion properties of the left Cayley graph $\Cay_L(A,G)$ and the right Cayley graph $\Cay_R(G,B)$ (i.e. the second largest magnitude of an eigenvalue of the adjacency matrix should be as small as possible), as well as the distance and robustness properties of the classical linear codes with parity check matrices $H_A \otimes H_B$ and $H_A^\bot \otimes H_B^\bot$. 

For the present comparison, we require a quantum Tanner code having length and dimension close to $[[512, 174]]$. Observing that $\Delta^3 \leq \Delta^2 |G|$, we see that the only possible choices are those with $\Delta \leq 8$; out of these, only the combination $\Delta=5, m=1$ gives rate around 174/512. Further, given $m=1$, we want the minimum distance of $\ker(H_A \otimes H_B)$, $\ker(H_A^\bot \otimes H_B^\bot)$ to be more than 1; this determines $H_A=H_B^\bot=\begin{pmatrix} 1 & 1 & 1 & 1 & 1 \end{pmatrix}$ and therefore a natural choice is 
\begin{align}
    H_A^\bot=H_B=
    \begin{pmatrix}
    1 &1 &0 &0 &0 \\
    0 &1 &1 &0 &0 \\
    0 &0 &1 &1 &0 \\
    0 &0 &0 &1 &1
    \end{pmatrix} .    
\end{align}
The last step is to choose $G,A,B$. Requiring a block size around 512 fixes $|G|=20$ or $21$. A computer search determines that the best spectral expansion properties are given by the choices 
\begin{align}
    G&=\langle s,t|s^4=t^5=1, ts=st^2 \rangle \\
    A&=\{1, s,s^3 ,t^2 s^2 ,t^3 s^2\} \\
    B&=\{1, t s^3,t^2 s, t^2 s^2 ,t^4 s^2 \} .
\end{align}
The resulting PCMs $H^x, H^z$ have rank 156 each, so we obtain a $[[500,188]]$ quantum Tanner code. The row weights of the parity check matrices are all 10, and the column weights are either 2 or 4. By exhaustive search of low-weight code words we have determined that the minimum distance of this code is 4.

\subsection{Decoding performance with the quantum erasure channel}

\begin{figure}
    \centering
    \begin{tabular}{cc} \hspace{-3mm}
    \scalebox{0.6}{\input{02_Fig_ErasureChannel}} & 
    \scalebox{0.6}{\input{03_Fig_DepolarisingChannel}}
    \end{tabular}
    \caption{Monte Carlo simulation of the logical error rates of 5 different CSS codes. Left: maximum likelihood decoding under erasure channel ($\mathcal{E}_\beta$). Right: BP decoding under depolarising channel ($\mathcal{D}_\epsilon$). See main text for details.}
    \label{fig:Simulations}
\end{figure}

The first error model we consider is the quantum erasure channel $\mathcal{E}_\beta$, which acts on a single qubit state $\rho$ as
\begin{align}
    \mathcal{E}_\beta(\rho) 
    = 
    \big(1-\beta\big) \, \rho \otimes | 0 \rangle \langle 0| 
    + \beta \, \frac{I}{2} \otimes |1 \rangle \langle 1 |   
\end{align}
where $\beta$ is a parameter of the channel specifying the probability of erasure (see~\cite[Chapter 5]{hayashi2006quantum} for an introduction to quantum channels). The rightmost qubit is interpreted as a ``flag'' which can be read-out deterministically (since $\ket{0}$ and $\ket{1}$ are orthogonal states) and which heralds whether an erasure has occurred. 

The main reason for considering the erasure channel is that it has been observed in~\cite[Section III]{delfosse2020linear} that for any stabilizer code on the erasure channel it is possible to efficiently perform maximum likelihood decoding: it is sufficient to find any error pattern, e.g.\ by Gaussian elimination, supported on the erased positions that produces the same syndrome\footnote{Technically, the decoder outputs an error pattern $e$ from the most probable coset of errors, that is, maximising the sum of the probabilities of all errors that differ from $e$ by a stabilizer of the code.}. This allows to test the performance of the codes in an unbiased manner, that is, in a way that is decoupled from the performance of efficient but sub-optimal decoding algorithms, such as belief propagation. We note, however, that  erasure errors do occur in certain quantum computation models or for concatenated codes where the inner decoder can signal a decoding failure. 

The probability of decoding error as a function of the probability of erasure is shown in Figure \ref{fig:Simulations} (left) for the bicycle, hypergraph product code (HPC), quantum Tanner code, our construction, and for a randomly chosen (not sparse) $[[512,174]]$ CSS code drawn from the probability distribution described in~\cite{ostrev2023qkdparameter}. In the present case, this means that a random $512 \times 512$ invertible matrix $A$ is chosen, then the first 169 rows of $A$ and the second 169 rows of $(A^{-1})^{\T}$ are chosen as $H^x$ and $H^z$, respectively. 

Our product code showcases the lowest logical error rate, compared to the other practically implementable codes, over the entire range of erasure probability $\beta$ that we have considered. This is mainly due to the fact that our code features a larger minimum distance: the minimum distances $d$ of the bicycle, HPC, quantum Tanner and our SPC(3) construction are 2, 4, 4 and 8, respectively. This results in an logical error rate scaling approximately as $\beta^d$. Furthermore, we have performed an exhaustive search to find the multiplicity of the minimum weight errors for the bicycle, HPC, quantum Tanner codes. The bicycle code has 28 weight-2 $X$ errors and, being self-dual, each for each $X$ error there exists a $Y$ error and a $Z$ error having the same support. The HPC has $861$ weight-4 $X$ errors, $861$ weight-4 $Z$ errors and no weight-4 $Y$ errors. The quantum Tanner code has $250$ weight-4 $X$ errors, $250$ weight-4 $Z$ errors and no weight-4 $Y$ errors. We remark that the lower number of weight-4 errors in the quantum Tanner code compared to the HPC results in a roughly constant offset in the logarithmic plots of their logical error rates.

\subsection{Decoding performance with the depolarising channel}

The second error model we consider is the depolarising channel $\mathcal{D}_\epsilon$, which acts on a single qubit state $\rho$ as
\begin{align}
   \mathcal{D}_\epsilon (\rho)
   & = 
   \big(1-\epsilon\big) \, \rho + 
   \frac{\epsilon}{3} \, \big(X \rho X + Y \rho Y + Z \rho Z\big)
   = 
   \left(1 - \frac{4\epsilon}{3} \right) \rho + 
   \frac{4\epsilon}{3} \frac{I}{2}
   \label{eq:depolarising}
\end{align}
where $\epsilon$ is a parameter of the channel. The first expression clarifies that $X$, $Y$ and $Z$ Pauli errors occur with equal probability, while the second expression is based on the identity $\frac{1}{4} (\rho + X\rho X+ Y\rho Y+ Z\rho Z) = I/2$ and shows that the completely depolarising channel is obtained for $\epsilon = 3/4$.

On the depolarising channel, there is no known efficiently computable maximum likelihood decoder. Instead, we use a quaternary belief propagation decoder: the messages are indexed by values in $\F_4$ or, equivalently, by a Pauli matrix. The advantage of using this decoder, instead of using two independent binary belief propagation decoders for $X$ and $Z$ checks is that it keeps into account the correlation between $X$ and $Z$ errors in the depolarising channel. Quaternary BP decoding was first introduced in~\cite{poulin2008iterative} (and see~\cite{babar2015fifteen} for a review of decoding algorithms for quantum LDPC codes).
For completeness we describe our implementation of the quaternary belief propagation decoder in Appendix~\ref{app:BP}. Further decoding improvement may be achieved using post-processing techniques such as ordered-statistics decoding~\cite{panteleev2021degenerate, roffe2020decoding, valls2021syndrome} and stabilizer inactivation~\cite{crest2022stabilizer}, or by exploiting the soft syndrome information that is accessible in some quantum computing hardware~\cite{fukui2018high, pattison2021improved, raveendran2022soft}.

The probability of decoding error as a function of the depolarising rate of the channel is shown in Figure~\ref{fig:Simulations} (right) for the bicycle code, the hypergraph product code, the quantum Tanner code, and the $[[512,174]]$ code from our 3-fold product construction; we also include the randomly chosen dense $[[512,174]]$ CSS code. We observe that our reference SPC$(3)$ code surpasses all other considered codes, judging by the BP error correction performance, over the considered range of depolarising noise intensities ($\epsilon$).

The decoding performance for the random CSS code is particularly poor. This is to be expected, since the code is non-sparse and therefore unsuitable for decoding via BP. In general, the maximum-likelihood decoding of quantum codes on the depolarising channel is a very hard problem (technically, \#P-hard~\cite{iyer2015hardness}). 

The decoding performance of the four sparse codes can be heuristically explained as follows. In general, it is known that the performance of the BP decoder is very sensitive to the structure of the Tanner graph. However, in the present case, it can be plainly seen that the two plots in Figure~\ref{fig:Simulations} are qualitatively the same for the sparse codes. Therefore, it is reasonable to conjecture that the dominant factor explaining the superior performance of the SPC code on the depolarizing channel is the same as on the erasure channel: its larger minimum distance. 

\section{Meta-checks and correction of syndrome readout errors}
\label{sec:syndrome_errors}

In this section, we show how meta-checks can be used to mitigate the effect of syndrome read-out errors. In particular, the SPC$(3,s)$ code has a distance with respect to syndrome errors that is equal to 3, allowing us to detect and locate one syndrome read-out error, which is a first step in the direction of fault-tolerance~\cite{gottesman1998theory, gottesman2010introduction}.

\subsection{Meta-check matrix}
Quantum read-out measurements are prone to errors, hence the syndrome outcomes may be faulty, i.e., incorrect. Fault-tolerant error correcting codes showcase the possibility of detecting and locating errors even in the face of syndrome read-out errors. The syndrome measurements are physically implemented via quantum circuits acting on both data and ancillary qubits and measuring out the ancillary qubits alone and thus correlated errors are typically present. Circuit-level noise simulations are therefore necessary to perform an accurate evaluation of the performance of a quantum error-correcting code~\cite{gidney2021stim}. Nonetheless, a simplified but informative analysis can be obtained using a so-called \textit{phenomenological error} model: data qubit errors and syndrome read-out errors are drawn independently at random, possibly with different error probabilities.

We now explain briefly how the presence of redundant checks helps to correct errors that occur during the syndrome read-out measurements. We start considering a generic CSS code and then we specialise to the SPC$(3,s)$ code. Let $H \in \F_2^{m \times n}$ be a matrix with rank $r<m$ (so that $m-r$ checks are redundant), $H$ can be either the $X$ or the $Z$ PCM of the CSS code. Consider the subspace $\mathcal{S}=\vspan(H) \subset \F_2^m$ of dimension $r$. If the device performing the syndrome measurement was perfect, the measured syndrome would be an element of $\mathcal{S}$. However, when the syndrome measurement is faulty, the measured syndrome is an element $s' \in \F_2^m$ that is not necessarily in $\mathcal{S}$; we may write $s' = He + e_s$, where $e_s$ represents a syndrome readout error. In that case, we regard $\mathcal{S}$ as a linear error-correcting code and try to correct the faulty $s'$ to the subspace $\mathcal{S}=\vspan(H)$. The faulty syndrome decoding works independently from any data-qubit error that may be present since the meta-checks, by construction, are insensitive to them. 

The matrix $H$ plays the role of a PCM for the quantum CSS code and of a generator matrix for $\mathcal{S}=\vspan(H)$. We now want to find another matrix $M$, such that $\ker(M)=\vspan(H)$; we call $M$ a \textit{meta-check} matrix. A systematic way for obtaining $M$ is the following: use Gaussian elimination to find an invertible $M'' \in \F_2^{m \times m}$ such that \begin{align}
\underbrace{
        \left(\begin{array}{c}
        ~M'~ \\ \hline  M 
        \end{array} \right)     
    }_{= M''}
    \bigg(
    ~H~
    \bigg) 
    =
    \left(\begin{array}{c}
    ~E~ \\ \hline  0 
    \end{array} \right)
\end{align}
where $M' \in\F_2^{r\times m}$ and $M \in \F_2^{(m-r)\times m}$ are full rank matrices, and where $E \in \F_2^{r \times n}$ is the reduced row echelon form of $H$. Then we have  $\ker(M) = \vspan(H)$ as needed. Therefore $Ms' = M(He + e_s) = Me_s$ and $M$ is insensitive to the data-qubit error $e$, as claimed.

The minimum distance $d_M$ of the meta-check matrix $M$ can be upper bounded as $d_M \leq w_c^{\min}$, where $w_c^{\min}$ is the minimum column weight of $H$. To see this, let $H^i$ a minimum-weight column of $H$; by assumption $H^i \in \vspan(H) = \ker(M)$, meaning that $H^i$ is a code word of $\mathcal{S}$ having weight $w_c^{\min}$, thus proving the claim. The proof that SPC($3,s$) codes have meta-check distance $d_M = w_c^{\min} = 3$ can be found in Appendix~\ref{app:proofs}, where furthermore we provide an explicit construction for a meta-check matrix $M$.

\subsection{Extended parity-check matrix}

Note that syndrome read-out errors $e_s$ do not directly affect the data qubits, so there is no physical correction to be applied. Nonetheless, a meta-syndrome $\sigma \in \F_2^{m-r}$ containing information about the location of the faulty syndrome can be obtained as  $\sigma = Ms$ and then forwarded to the decoder that is tasked with decoding the quantum CSS code associated to $H$, thus yielding a two-step decoding. However, as noted in~\cite[Section VII]{higgott2023improved} the decoding performance typically improves if one uses a single stage BP decoder directly. To this end we define \textit{extended PCM, extended error} and \textit{extended syndrome} $\widetilde{H}, \widetilde{e}, \widetilde{s}$ via 
\begin{align}
\underbrace{
    \left(\begin{array}{c|c}
    ~H~ &  ~I_m~ \\ \hline
    0   &   M  \\     
    \end{array} \right) }_{=\widetilde{H}}
\underbrace{
    \left(\begin{array}{c}
    e   \\ \hline
    e_s \\     
    \end{array}\right) }_{=\widetilde{e}}
    =
    \left(\begin{array}{c}
    He + e_s  \\ \hline
    Me_s \\     
    \end{array}\right)
    =
\underbrace{
    \left(\begin{array}{c}
    s' \\ \hline
    \sigma \\     
    \end{array}\right)}_{=\widetilde{s}}
\label{eq:H_ext}
\end{align}
so that  $\widetilde{s} = \widetilde{H} \widetilde{e}$ holds. The extended PCM $\widetilde{H}$ can be interpreted as adding one hidden variable for each syndrome which may flip the result of the measurements and adding a meta-check code $M$ to correct the effect of these hidden variables. The single-stage BP decoder takes as input the extended syndrome $\widetilde{s} \in \F_2^{m+(m-r)}$ and aims at reconstructing the $\widetilde{e} \in \F_2^{n+m}$ employing $\widetilde{H}$ to specify the bipartite graph for the massage passing algorithm. It is also possible to extend the quaternary BP that simultaneously decodes the $X$ and $Z$ components of the CSS code to include the meta-checks, see Appendix~\ref{app:BP} for further details.

A standard way for obtaining resilience to readout errors is to repeat all the measurements $k$ times. This can be interpreted as protecting the readout data via a repetition code; a code with PCM $H$ would be modified to $H_{(k)} = 1_k \otimes H$, where $1_k = (1,\cdots,1)^\T \in \F_2^{k}$, while the associated meta-check matrix is $M=H_{k\text{-rep}} \otimes I$, where $H_{k\text{-rep}}$ is the PCM of the $k$-repetition code. Doing so results in a large overhead in the number of measurements, scaling linearly in the number of repetitions. Product codes can increase the meta-check distance much more efficiently. For instance, the SPC(3) code requires a minimum of $338$ measurements to be stabilized, while the associated PCM defined in Eqs.~\eqref{3D-SPC_x} and~\eqref{3D-SPC_z} achieve a meta-check distance equal to $3$ while requiring only $384$ measurements, an increase of about $13.6\%$ compared to the minimum.

\subsection{Decoding performance including syndrome readout errors}

\begin{figure}
	\centering
	\begin{tabular}{cc} \hspace{-3mm}
		\scalebox{0.6}{\input{04_Fig_DepolarisingChannel_SPAM}} & 
		\scalebox{0.6}{\input{05_Fig_DepolarisingChannel_SPAM_only_CSS_product}}
	\end{tabular}
	\caption{Monte Carlo simulations of the logical error rates under BP decoding under a depolarizing channel with syndrome measurement errors. Left: all codes are subject to either no syndrome readout error (solid lines) or to a syndrome readout flip with probability $p=10^{-3}$ (dashed lines). Right: performance of the SPC(3) code for different values of the syndrome readout error probability $p$.}
	\label{fig:Simulations_meta}
\end{figure}

We have benchmarked the resilience to syndrome readout errors of the SPC(3) code under extended quaternary BP decoding. The employed phenomenological error model is given by the depolarising channel of Eq.~\eqref{eq:depolarising} acting on the data qubits, while each syndrome is subject to readout errors that may flip the outcome with probability $p$ (i.e., it is a classical binary symmetric channel). The BP messages are passed along the edges of a bipartite graph implicitly defined by $\widetilde{H}$ as in Eq.~\eqref{eq:H_ext}, where the meta-check matrix $M$ is the one provided in Appendix~\ref{app:proofs}. The results are presented in Figure~\ref{fig:Simulations_meta}.

As a preliminary step, we have verified that by fixing $p=0$ and running the quaternary BP decoding over the graphs defined by $\widetilde{H}$ and $H$, respectively, yield results that are essentially indistiguishable (blue solid line in Figure~\ref{fig:Simulations}, right, and in Figure~\ref{fig:Simulations_meta}, left). This shows that BP is neither impaired nor enhanced by the extension of the decoding graph for the SPC(3) code.

We have then assessed the decoding performance of the codes described in Section~\ref{sec:previous_works} in presence of readout errors, with the exclusion of the random CSS code, which cannot be decoded via BP. The simulation results are presented as dashed lines in Figure~\ref{fig:Simulations_meta}, left, for $p=10^{-3}$. This readout fidelity is comparable with what can be obtained, e.g., in state of the art quantum computers based on Rydberg atom arrays~\cite{bluvstein2024logical}. As expected, the codes that do not feature meta-checks, are significantly affected by readout errors. In contrast, the SPC(3) logical error rate is essentially unaffected for depolarisation rates $\epsilon > 6\cdot 10^{-4}$, below which it converges to an error floor of around $3\cdot 10^{-5}$.

When increasing $p$ to higher values, $p=10^{-2}$ or $p=10^{-1}$, the performance of SPC(3) with quaternary BP decoding significantly degrades, see Figure~\ref{fig:Simulations_meta}, right. This means that repeating the syndrome measurements a few times may be required in order to increase the meta-check distance and thus counter the effect of readout errors.

\section{Conclusions and outlook}\label{sec:conclusions}

In this work, we have introduced new methods for constructing families of CSS codes associated to classical product codes. Quantum CSS codes are challenging to construct (compared to classical ones) since a commutativity condition between $X$ and $Z$ checks must be satisfied, as given in Eq.~\eqref{eq:CSS}. Our first construction for CSS product codes is asymmetrical, in the sense that $X$ parity checks are constructed from a classical product code and $Z$ parity checks from a classical tensor product code, with the goal of automatically satisfying Eq.~\eqref{eq:CSS}. A second construction is more symmetrical, having both $X$ and $Z$ checks associated to classical product codes, while the component codes are in the form of classical tensor product codes; this second construction showcases an increased (pure) distance against both kinds of Pauli errors compared to the component CSS codes it is constructed from. The third construction is a multi-dimensional generalisation of the second one, allowing to obtain $D$-fold products of component CSS codes.

An advantage of our classically-inspired product code construction is that extensive parallelisation can be achieved in measuring the parity checks. Using the array representation of a 2-fold product code (see Figure~\ref{fig:ProductCode}) it is clear that the parity checks of the component codes can be applied in parallel over all the rows of the array and, in a second step, measured in parallel over all the columns of the array. The same is true for CSS product codes and, if the qubits are physically arranged in a 2-dimensional array, as in the quantum computing architectures based on Rydberg atom~\cite{barredo2016atom, gross2017quantum, arute2019quantum, holz2020-2d, wu2021strong, xu2024constant, bluvstein2024logical}, this can result in simplified hardware implementations. A caveat is that the columns of the PCM associated to $Z$ are permuted compared to the PCM associated to $Z$ checks: this means that the $Z$ syndrome measurements act on qubits that are no longer aligned along single rows or columns of the 2-dimensional array. 

Product codes yield good performances when using SPC codes as the component codes (albeit other choices are possible~\cite{chiaraluce2004extended, le2008reed, mukhtar2016turbo}). The heuristic motivation is that SPC codes have high encoding rate (only one parity bit of overhead) but small distance ($d=2$); the product construction allows to increase the distance at the expense of decreasing the rate and yields codes having good properties for intermediate code sizes. These good properties are retained also by the corresponding quantum CSS codes. For instance, we obtain a 3-fold product SPC code that can encode 174 logical qubits in 512 physical qubits ($R = 174/512 \simeq 0.34$) and having distance $d=8$. This code has good performances both for erasure channels and for depolarising noise under BP decoding; for instance, the second best code in our comparison is a $[[500, 188]]$ quantum Tanner code, which exhibits logical error rates that are around two orders of magnitude higher for realistic values of the depolarising noise. 

Given the promising results of these product codes, we aim to further investigate their properties. Open questions regard whether there are other good choices for the component codes, besides SPC codes. This could be especially true for the asymmetric product construction, where the fact that only two component CSS codes are needed (instead of the four required in the symmetric construction) gives leeway in the choice of larger component CSS codes. Another interesting question is the possibility of achieving better practical results via code concatenation, e.g., using a surface code for low-level error correction and a product code as a high-level code to further boost the fidelity of the encoded logical qubits~\cite{chamberland2022building, li2023concatenation, raveendran2022finite, cao2022quantum}.

Regarding the code minimum distance $d$, it would be interesting to find better lower bounds, since bounds based on the pure distance $\delta$ are in general not tight (but we have shown that for the special case of SPC product codes we have $d = \delta$). The fault-tolerance properties of these codes could be further investigated, simulating the performance of decoders under realistic circuit-level noise~\cite{gidney2021stim}. Finally, methods for implementing fault-tolerant Clifford gates~\cite{rengaswamy2018synthesis} and magic-state distillation~\cite{bravyi2005universal, litinski2019magic} within these code families should be ascertained for realising universal quantum computation with CSS product codes.

\section*{Acknowledgements}

We acknowledge Dr.\ Francesco Guatieri for optimising the code for the exhaustive search of the minimum weight errors.  This project was funded within the QuantERA II Programme that has received funding from the European Union’s Horizon 2020 research and innovation programme under Grant Agreement No 101017733, through the project EQUIP. 

\appendix

\input{Appendices}

\end{document}

%% file: 01_Fig_ProductCode.tex
\begin{tikzpicture}[scale=0.8]

  \draw[pattern=north west lines, pattern color=blue!25!] (6,0) rectangle (9,6);
  \draw[pattern=north east lines, pattern color= red!25!] (0,0) rectangle (9,2);

  \draw[step=1,black,thin] (0,0) grid (9,6);
  
  \draw[black, line width=1.2pt] (0,0) rectangle (9,2);
  \draw[black, line width=1.2pt] (6,0) rectangle (9,6);
  
  \foreach \x in {1, ..., 9} {%
    \foreach[evaluate={\v = int(\x + 9*(\y-1))}] \y in {1, ..., 6} {%
      \draw node[xshift=-11, yshift=11] at (\x,6-\y) {\large $b_{\v}$};
    };
  };

  \draw [
    line width=1.2pt,
    decorate, 
    decoration = {brace, raise=4pt, amplitude=6pt}] 
  (0,6)  --  (5.95,6)
  node[pos=0.5, above=12pt]{\large $k_1 = 6$};  

  \draw [
    line width=1.2pt,
    decorate, 
    decoration = {brace, raise=4pt, amplitude=6pt}] 
  (6.05,6)  --  (9,6)
  node[pos=0.5, above=12pt]{\large $m_1 = 3$};  
  
  \draw [
    line width=1.2pt,
    decorate, 
    decoration = {brace, raise=4pt, amplitude=6pt}] 
  (0,2.05)  --  (0,6)
  node[pos=0.5,left=12pt]{\large $k_2 = 4$};

  \draw [
    line width=1.2pt,
    decorate, 
    decoration = {brace, raise=4pt, amplitude=6pt}] 
  (0,0)  --  (0,1.95)
  node[pos=0.5,left=12pt]{\large $m_2 = 2$};

\end{tikzpicture}

%% file: 02_Fig_ErasureChannel.tex
\begin{tikzpicture}

\begin{axis}[%
width=10cm,
height=8cm,
scale only axis,
xmin=0.01,
xmax=0.4,
xmode=log,
xlabel={\large{Qubit erasure probability $\beta$}},
ymin=1E-5,
ymax=1,
ymode=log,
ylabel={\large{Logical error rate}},
axis background/.style={fill=white},
xminorgrids,
yminorgrids,
legend style={at={(0.01,0.99)}, anchor=north west, legend cell align=left, align=left, draw=white!15!black}
]
\addplot [color=green, solid, mark=diamond*,  mark options={solid, fill=green!50}, mark size= 3, each nth point=1, filter discard warning=false, unbounded coords=discard]
table[row sep=crcr]{%
0.01	0.0023\\
0.0144612554959193	0.0049\\
0.0209127910518255	0.0093\\
0.0302425214533222	0.0196\\
0.0437344829577311	0.0414\\
0.0632455532033676	0.0816\\
0.0914610103854653	0.174\\
0.132264103909914	0.344\\
0.191270499958007	0.692\\
0.276601156872496	0.974\\
0.4	1\\
};
\addlegendentry{bicycle }

\addplot [color=red, solid, mark=*,  mark options={fill=red!60}, mark size= 2.5]
table[row sep=crcr,each nth point=1, filter discard warning=false, unbounded coords=discard]{%
0.01	1e-05\\
0.0144612554959193	5e-05\\
0.0209127910518255	1.45e-04\\
0.0302425214533222	7.6563e-04\\
0.0437344829577311	0.0033\\
0.0632455532033676	0.0141\\
0.0914610103854653	0.0631\\
0.132264103909914	0.2458\\
0.191270499958007	0.7205\\
0.276601156872496	0.9920\\
0.4	1\\
};
\addlegendentry{HPC}

\addplot [color=orange, solid, mark=pentagon*, mark options={orange!60}, mark size= 3, filter discard warning=false, unbounded coords=discard]
table[row sep=crcr]{%
0.0144612554959193	7.8125e-06\\
0.0209127910518255	0.00003515625\\
0.0302425214533222	0.00021875\\
0.0437344829577311	0.001\\
0.0632455532033676	0.00425\\
0.0914610103854653	0.01775\\
0.132264103909914	0.0895\\
0.191270499958007	0.426\\
0.276601156872496	0.974\\
0.4	1\\
};
\addlegendentry{q. Tanner}

\addplot [color=blue, solid, mark=square*,  mark options={fill=blue!60},mark size= 2.5, filter discard warning=false, unbounded coords=discard ]
table[row sep=crcr]{%
0.0632455532033676	9.14e-06\\
0.0914610103854653	0.00011\\
0.132264103909914	0.002\\
0.191270499958007	0.046\\
0.276601156872496	0.694\\
0.4	1\\
};
\addlegendentry{SPC(3)}

\addplot [color=violet, solid, mark=triangle*,  mark options={fill=violet!60}, mark size= 3, each nth point=1, filter discard warning=false, unbounded coords=discard]
table[row sep=crcr]{%
0.26    3e-04\\
0.276601156872496   0.0043\\
0.29    0.0346\\
0.31    0.1892\\
0.33    0.522\\
0.35    0.835\\
0.37    0.9744\\
0.40    0.9994\\
};
\addlegendentry{random CSS}

\end{axis}
\end{tikzpicture}%

%% file: 03_Fig_DepolarisingChannel.tex
\begin{tikzpicture}

\begin{axis}[%
width=10cm,
height=8cm,
scale only axis,
xmode=log,
xmin=0.001,
xmax=0.1,
xminorticks=true,
xlabel={\large{Depolarising rate $\epsilon$}},
ymode=log,
ymin=1e-05, 
ymax=1,
yminorticks=true,
ylabel={\large{Logical error rate}},
axis background/.style={fill=white},
xminorgrids,
yminorgrids,
legend style={at={(0.98,0.02)}, anchor=south east, legend cell align=left, align=left, draw=white!15!black}
]
\addplot  [color=green, solid, mark=diamond*,  mark options={fill=green!60},mark size= 3]
  table[row sep=crcr]{%
0.1	1\\
0.0630957344480193	0.998003992015968\\
0.0398107170553497	0.961538461538462\\
0.0251188643150958	0.871080139372822\\
0.0158489319246111	0.67476383265857\\
0.01	0.488758553274682\\
0.00630957344480193	0.412881915772089\\
0.00398107170553497	0.248262164846077\\
0.00251188643150958	0.17082336863683\\
0.00158489319246111	0.111482720178372\\
0.001	0.0791765637371338\\
};
\addlegendentry{bicylcle }

\addplot [color=red, solid, mark=*,  mark options={fill=red!60},, mark size= 3]
  table[row sep=crcr]{%
0.1	1\\
0.0630957344480193	0.985221674876847\\
0.0398107170553497	0.896860986547085\\
0.0251188643150958	0.579710144927536\\
0.0158489319246111	0.325732899022801\\
0.01	0.142247510668563\\
0.00630957344480193	0.0559127760693318\\
0.00398107170553497	0.0209709552270106\\
0.00251188643150958	0.00868960722975321\\
0.00158489319246111	0.00315870935135903\\
0.001	0.00141172152381221\\
};
\addlegendentry{HPC}

\addplot  [color=orange, solid, mark=pentagon*,  mark options={fill=orange!60}, mark size= 3]
  table[row sep=crcr]{%
0.1	0.99009900990099\\
0.0630957344480193	0.847457627118644\\
0.0398107170553497	0.464037122969838\\
0.0251188643150958	0.195503421309873\\
0.0158489319246111	0.074487895716946\\
0.01	0.0266347050206419\\
0.00630957344480193	0.0102917717285031\\
0.00398107170553497	0.00417336143396699\\
0.00251188643150958	0.00173143683285575\\
0.00158489319246111	0.000654662341530797\\
0.001	0.00030429839744292\\
};
\addlegendentry{q. Tanner}

\addplot [color=blue, solid, mark=square*,  mark options={fill=blue!60}, mark size= 2.5]
  table[row sep=crcr]{%
0.1	0.956937799043062\\
0.0630957344480193	0.503778337531486\\
0.0398107170553497	0.0770416024653313\\
0.0251188643150958	0.0119196614816139\\
0.0158489319246111	0.00193457274960825\\
0.01	0.000303480926223787\\
0.00630957344480193	7.68609968871296e-05\\
0.00398107170553497	1.4376e-05\\
};
\addlegendentry{SPC(3)}

\addplot [color=violet, solid, mark=triangle*, mark options={fill=violet!60}, mark size= 3]
  table[row sep=crcr]{%
0.1	1\\
0.0630957344480193	1\\
0.0398107170553497	1\\
0.0251188643150958	1\\
0.0158489319246111	1\\
0.01	0.930232558139535\\
0.00630957344480193	0.873362445414847\\
0.00398107170553497	0.784313725490196\\
0.00251188643150958	0.66006600660066\\
0.00158489319246111	0.515463917525773\\
0.001	0.383877159309021\\
};
\addlegendentry{random CSS}

\end{axis}
\end{tikzpicture}

%% file: 04_Fig_DepolarisingChannel_SPAM.tex
\begin{tikzpicture}

\begin{axis}[%
width=10cm,
height=8cm,
scale only axis,
xmode=log,
xmin=0.001,
xmax=0.1,
xminorticks=true,
xlabel={\large{Depolarising rate $\epsilon$}},
ymode=log,
ymin=1e-05, 
ymax=1,
yminorticks=true,
ylabel={\large{Logical error rate}},
axis background/.style={fill=white},
xminorgrids,
yminorgrids,
legend style={at={(0.99,0.01)}, anchor=south east, legend cell align=left, align=left, draw=white!15!black,nodes={scale=0.95, transform shape} }
]
\addplot  [color=green, solid, mark=diamond*, mark options={fill=green!60}, mark size= 3]
  table[row sep=crcr]{%
0.1	1\\
0.0630957344480193	0.998003992015968\\
0.0398107170553497	0.961538461538462\\
0.0251188643150958	0.871080139372822\\
0.0158489319246111	0.67476383265857\\
0.01	0.488758553274682\\
0.00630957344480193	0.412881915772089\\
0.00398107170553497	0.248262164846077\\
0.00251188643150958	0.17082336863683\\
0.00158489319246111	0.111482720178372\\
0.001	0.0791765637371338\\
};
\addlegendentry{bicylcle, $p=0$}

\addplot  [color=green, dashed, mark=diamond, mark options={solid}, mark size= 3]
table[row sep=crcr]{%
0.1	1\\
0.0630957344480193	0.998003992015968\\
0.0398107170553497	0.959692898272553\\
0.0251188643150958	0.881834215167548\\
0.0158489319246111	0.755287009063444\\
0.01	0.559284116331096\\
0.00630957344480193	0.424088210347752\\
0.00398107170553497	0.314267756128221\\
0.00251188643150958	0.212675457252233\\
0.00158489319246111	0.132205182443152\\
0.001	0.0899604174163368\\
};
\addlegendentry{bicylcle, $p=10^{-3}$}

\addplot [color=red, solid, mark=*, mark options={fill=red!60}, mark size= 3]
  table[row sep=crcr]{%
0.1	1\\
0.0630957344480193	0.985221674876847\\
0.0398107170553497	0.896860986547085\\
0.0251188643150958	0.579710144927536\\
0.0158489319246111	0.325732899022801\\
0.01	0.142247510668563\\
0.00630957344480193	0.0559127760693318\\
0.00398107170553497	0.0209709552270106\\
0.00251188643150958	0.00868960722975321\\
0.00158489319246111	0.00315870935135903\\
0.001	0.00141172152381221\\
};
\addlegendentry{HPC, $p=0$}

\addplot [color=red, dashed, mark=o, mark options={solid}, mark size= 3]
table[row sep=crcr]{%
0.630957344480193	1\\
0.398107170553497	1\\
0.251188643150958	1\\
0.158489319246111	1\\
0.1	1\\
0.0630957344480193	0.985221674876847\\
0.0398107170553497	0.904977375565611\\
0.0251188643150958	0.666666666666667\\
0.0158489319246111	0.37593984962406\\
0.01	0.161420500403551\\
0.00630957344480193	0.0759878419452888\\
0.00398107170553497	0.0353045013239188\\
0.00251188643150958	0.015500271254747\\
0.00158489319246111	0.00920047842487809\\
0.001	0.00417597561230242\\
};
\addlegendentry{HPC, $p=10^{-3}$}

\addplot  [color=orange, solid, mark=pentagon*, mark options={fill=orange!60},mark size= 3]
  table[row sep=crcr]{%
0.1	0.99009900990099\\
0.0630957344480193	0.847457627118644\\
0.0398107170553497	0.464037122969838\\
0.0251188643150958	0.195503421309873\\
0.0158489319246111	0.074487895716946\\
0.01	0.0266347050206419\\
0.00630957344480193	0.0102917717285031\\
0.00398107170553497	0.00417336143396699\\
0.00251188643150958	0.00173143683285575\\
0.00158489319246111	0.000654662341530797\\
0.001	0.00030429839744292\\
};
\addlegendentry{q. Tanner, $p=0$ }

\addplot  [color=orange, dashed, mark=pentagon, mark options={solid}, mark size= 3]
table[row sep=crcr]{%
0.630957344480193	1\\
0.398107170553497	1\\
0.251188643150958	1\\
0.158489319246111	1\\
0.1	1\\
0.0630957344480193	0.900900900900901\\
0.0398107170553497	0.550964187327824\\
0.0251188643150958	0.258397932816537\\
0.0158489319246111	0.11154489682097\\
0.01	0.0635324015247776\\
0.00630957344480193	0.028534741047225\\
0.00398107170553497	0.0164028540966128\\
0.00251188643150958	0.00968757568418503\\
0.00158489319246111	0.00504044960810504\\
0.001	0.00274040174289551\\
};
\addlegendentry{q. Tanner, $p=10^{-3}$ \hspace{-3.5mm} }

\addplot [color=blue, solid, mark=square*,mark options={fill=blue!60}, mark size= 2.5]
  table[row sep=crcr]{%
0.1	0.956937799043062\\
0.0630957344480193	0.503778337531486\\
0.0398107170553497	0.0770416024653313\\
0.0251188643150958	0.0119196614816139\\
0.0158489319246111	0.00193457274960825\\
0.01	0.000303480926223787\\
0.00630957344480193	7.68609968871296e-05\\
0.00398107170553497	1.4376e-05\\
};
\addlegendentry{ SPC(3), $p=0$}

\addplot [color=blue, dashed, mark=square, mark options={solid}, mark size= 2.5]
table[row sep=crcr]{%
0.630957344480193	1\\
0.398107170553497	1\\
0.251188643150958	1\\
0.158489319246111	1\\
0.1	0.980392156862745\\
0.0630957344480193	0.50761421319797\\
0.0398107170553497	0.0922509225092251\\
0.0251188643150958	0.0126246686024492\\
0.0158489319246111	0.0019018999980981\\
0.01	0.000318527694213414\\
0.00630957344480193	9.135724e-05\\
0.00398107170553497	5.128205e-05\\
0.00251188643150958	3.453035e-5\\
};
\addlegendentry{ SPC(3), $p=10^{-3}$ }
\end{axis}
\end{tikzpicture}

%% file: 05_Fig_DepolarisingChannel_SPAM_only_CSS_product.tex
\begin{tikzpicture}

\begin{axis}[%
width=10cm,
height=8cm,
scale only axis,
xmode=log,
xmin=0.001,
xmax=0.1,
xminorticks=true,
xlabel={\large{Depolarising rate $\epsilon$}},
ymode=log,
ymin=1e-05, 
ymax=1,
yminorticks=true,
ylabel={\large{Logical error rate}},
axis background/.style={fill=white},
xminorgrids,
yminorgrids,
legend style={at={(0.98,0.02)}, anchor=south east, legend cell align=left, align=left, draw=white!15!black}
]
\addplot  [color=blue, solid, mark=square*,mark options={fill=blue!60}, mark size= 2.5]
  table[row sep=crcr]{%
0.1	0.956937799043062\\
0.0630957344480193	0.503778337531486\\
0.0398107170553497	0.0770416024653313\\
0.0251188643150958	0.0119196614816139\\
0.0158489319246111	0.00193457274960825\\
0.01	0.000303480926223787\\
0.00630957344480193	7.68609968871296e-05\\
0.00398107170553497	1.4376e-05\\
};
\addlegendentry{$p=0$}

\addplot   [color=blue, dashed, mark=square, mark options={solid}, mark size= 2.5]
  table[row sep=crcr]{%
0.630957344480193	1\\
0.398107170553497	1\\
0.251188643150958	1\\
0.158489319246111	1\\
0.1	0.980392156862745\\
0.0630957344480193	0.50761421319797\\
0.0398107170553497	0.0922509225092251\\
0.0251188643150958	0.0126246686024492\\
0.0158489319246111	0.0019018999980981\\
0.01	0.000318527694213414\\
0.00630957344480193	9.135724e-05\\
0.00398107170553497	5.128205e-05\\
0.00251188643150958	3.453035e-5\\
};
\addlegendentry{$p=10^{-3}$}

\addplot  [color=blue, dashed, mark=10-pointed star, mark options={solid}, mark size= 3]
  table[row sep=crcr]{%
0.630957344480193	1\\
0.398107170553497	1\\
0.251188643150958	1\\
0.158489319246111	1\\
0.1	0.995024875621891\\
0.0630957344480193	0.5\\
0.0398107170553497	0.106382978723404\\
0.0251188643150958	0.0234439104442621\\
0.0158489319246111	0.00681059728938228\\
0.01	0.0029556941447699\\
0.00630957344480193	0.00204492704722759\\
0.00398107170553497	0.00105987217941516\\
0.00251188643150958	0.000547345374931582\\
0.00158489319246111	0.0003638156255173\\
0.001	0.000243638002655654\\
};
\addlegendentry{$p=10^{-2}$}

\addplot  [color=blue, dashed, mark=+, mark options={solid}, mark size= 3]
  table[row sep=crcr]{%
0.630957344480193	1\\
0.398107170553497	1\\
0.251188643150958	1\\
0.158489319246111	1\\
0.1	1\\
0.0630957344480193	1\\
0.0398107170553497	0.938967136150235\\
0.0251188643150958	0.760456273764259\\
0.0158489319246111	0.481927710843373\\
0.01	0.258397932816537\\
0.00630957344480193	0.132362673726009\\
0.00398107170553497	0.0785545954438335\\
0.00251188643150958	0.0562587904360056\\
0.00158489319246111	0.0291417747340813\\
0.001	0.0180831826401447\\
};
\addlegendentry{$p=10^{-1}$}

\end{axis}
\end{tikzpicture}

%% file: Appendices.tex
\section{Omitted proofs of code distance properties}
\label{app:proofs}

\subsection{Classical product code minimum distance: $d_\times = d_1 d_2$}

The codes we consider are $\F_2$-linear, hence the code minimum distance is equal to the minimum Hamming weight of any non-zero code-word. Consider then minimum distance code-words $x_1 \in \cC_1$ and $x_2 \in \cC_2$, and denote their Hamming weights $d_1$ and $d_2$, respectively. The non-zero code-word $x = x_1 \otimes x_2 \in \cC_1 \times \cC_2$ has weight $d_1 d_2$, thus the product code has distance $d_\times \leq d_1 d_2$. Conversely, all non-zero code-words must have at least $d_2$ rows that are not identically zero and each of these rows must have at least $d_1$ ones in it. Therefore, $d_\times \geq d_1 d_2$.

\subsection{Classical tensor product code minimum distance: $d_\otimes = \min (d_1,d_2)$}

Given any minimum weight words $x \in \cC_1, y \in \cC_2$ (i.e., $|x| = d_1$, $|y| = d_2$, $H_1 x = 0$, $H_2 y = 0$) and given any two unit vectors $\e_i$ and $\e_j$ (having a one in position $i$ and in position $j$, respectively, and zeros elsewhere) we have $|x \otimes \e_i| = d_1$, $|\e_j \otimes y| = d_2$, and moreover 
\begin{align}
    H_\otimes (x \otimes \e_i) 
    =   
    H_1 x \otimes H_2 \e_i
    = 
    0 \otimes H_2 \e_i
    = 0\,.
\end{align}
Analogously we have $H_\otimes (\e_j \otimes y) = 0$, which proves $d_\otimes \leq \min (d_1,d_2)$. Conversely, assume by contradiction that $w \in \cC_\otimes$ is a non-zero code-word having Hamming weight $d < \min (d_1,d_2)$. We can write $w$ as the sum of $d$ unit vectors, $w = \sum_{\alpha=1}^d \e_{i(\alpha)} \otimes \e_{j(\alpha)}$. The vectors $x_\alpha := H_1 \e_{i(\alpha)}$ are linearly independent, since otherwise we could find a word of $\cC_1$ with Hamming weight less or equal to $d < d_1$. Similarly, $y_\alpha := H_2 \e_{j(\alpha)}$ are linearly independent. By the definition of tensor product, $\{ x_\alpha \otimes y_\beta \}_{\alpha,\beta}$ are linearly independent, which implies
\begin{align}
    0 
    \neq 
    \sum_{\alpha=1}^d 
    x_\alpha \otimes y_\alpha
    =
    \sum_{\alpha=1}^d 
    H_1 \e_{i(\alpha)} \otimes H_2 \e_{j(\alpha)}
    =
    (H_1 \otimes H_2) w
\end{align}
contra the hypothesis that $w \in \cC_\otimes$, hence we must have $d_\otimes \geq \min (d_1,d_2)$.

\subsection{Examples of asymmetric CSS product codes having $d_\ltimes^z < d_1^z d_2^z$}

Let us construct an asymmetric product CSS codes where the two base codes are both the 9-qubit Shor code. The $X$ and $Z$ PCMs of the Shor code are given by, respectively,
\begin{align}
    H^x := 
    \left(
    \begin{array}{ccccccccc}
     1 & 1 & 1 & 1 & 1 & 1 & 0 & 0 & 0 \\
     0 & 0 & 0 & 1 & 1 & 1 & 1 & 1 & 1 \\
    \end{array}
    \right)
    \qquad
    H^z := 
    \left(
    \begin{array}{ccccccccc}
     1 & 1 & 0 & 0 & 0 & 0 & 0 & 0 & 0 \\
     0 & 1 & 1 & 0 & 0 & 0 & 0 & 0 & 0 \\
     0 & 0 & 0 & 1 & 1 & 0 & 0 & 0 & 0 \\
     0 & 0 & 0 & 0 & 1 & 1 & 0 & 0 & 0 \\
     0 & 0 & 0 & 0 & 0 & 0 & 1 & 1 & 0 \\
     0 & 0 & 0 & 0 & 0 & 0 & 0 & 1 & 1 \\
    \end{array}
    \right)    
\end{align}
and note that it is a CSS code having distances $d^x=3, d^z=3$ and pure distances $\delta^x=3,\delta^z=2$; for instance, $Z^{\otimes 2} \otimes I^{\otimes 7}$ (corresponding to the vector $\e_1 + \e_2 \in \F_2^{9}$) is a weight 2 undetectable Pauli operator which acts as the identity on the logical subspace. We define the asymmetric product CSS code $\cC_\ltimes^\text{Shor}$ via the PCMs:
\begin{align}
\label{eq:Shor_asym}
  H^x_\ltimes
    = 
    \begin{pmatrix}
	  H^x \o I \\
	  I \o H^x        
    \end{pmatrix}
  \qquad
  H^z_\ltimes
  = 
  \Big(\!\!
    \begin{array}{ccc}
	  H^z \o H^z
    \end{array}
  \!\!\Big) .
\end{align}
The weight-$6$ Pauli-$Z$ operator associated to the vector 
\begin{align}
    v^z :=  (\e_1 + \e_2) \otimes (\e_1 + \e_4 + \e_7)
\end{align}
is as undetectable logical error for $\cC_\ltimes^\text{Shor}$. First, notice that $H^x (\e_1 + \e_2) = H^x (\e_1 + \e_4 + \e_7) = 0$, hence $v^z$ is undetected by $X$ parity checks, $H^x_\ltimes v^z = 0$. Second, let $u=e_1 \otimes \sum_{i=1}^9 e_i$, and note that $H_\ltimes^z u =0 $ while $u^\T v^z=1$; therefore $v^z$ is not a stabiliser, $v^z \notin \vspan \!\big((H_\ltimes^z)^\T\big)$. This shows that $d_\ltimes^x \leq 6 < 9 = d_1^x d_2^x$.

The pattern can be extended as follows. Let 
\begin{equation}
H_D = \begin{pmatrix} 
1 & 1 & 0 & 0 & \cdots & 0 & 0\\
0 & 1 & 1 & 0 & \cdots & 0 & 0\\
\vdots & \vdots & \vdots & \vdots & \vdots & \vdots & \vdots\\
0 & 0 & 0 & 0 & \cdots & 1 & 1
\end{pmatrix} \in \F_2^{(D-1) \times D}
\;\;\;\;
1_D = \begin{pmatrix}
      1 \\ 1 \\ \vdots \\ 1
\end{pmatrix} \in \F_2^D
\end{equation}
The generalised code Shor$(D)$ has parity check matrices
$H^x = H_D \otimes 1_D^\T$, $H^z = I_D \otimes H_D$ and distances $d^x = d^z = \delta^x = D$ and $\delta^z = 2$. We then consider the asymmetric product CSS code $\cC_\ltimes^{\text{Shor}(D)}$ using again Eq.~\eqref{eq:Shor_asym}. Let us consider the weight-$2D$ Pauli-$Z$ operator associated to the vector $v^z :=  e_1 \otimes (e_1 + e_2) \otimes 1_D \otimes e_1$ as well as the vector $u=e_1\otimes e_1 \otimes e_1 \otimes 1_D$. From $H^x_\ltimes v^z=0$, $H^z_\ltimes u =0$, $u^\T v^z = 1$, we see that $v^z$ is an undetected logical error, therefore the distance of $\cC_\ltimes^{\text{Shor}(D)}$ is upper bounded as $d_\ltimes^z \leq 2 D$.

\subsection{Single-parity-check D-fold product CSS code distance: $d_{\mathrm{SPC}(D,s)} = 2^D$}

We have shown in the main text that $\delta_{\mathrm{SPC}(D,s)} = 2^D$, thus $d_{\mathrm{SPC}(D,s)} \geq 2^D$. Here we show that there exist logical errors of weight $2^D$, thus establishing $d_{\mathrm{SPC}(D,s)} \leq 2^D$.

We claim that the following vector corresponds to undetectable logical error of weight $2^D$ for both $H_{\mathrm{SPC}(D)}^x$ and $H_{\mathrm{SPC}(D)}^z$:
\begin{align}
\label{w_D}
    w_{(D)} = \bigotimes_{\ell=1}^{D^2} a_\ell
    \qquad
    a_\ell :=
    \begin{cases}
    u \, := (1, 1, \underbrace{0 ~ \cdots ~ 0}_{2(s-1)})^\T & \text{for } \ell=(i-1)D+i, \; i=1,\dots D \\
    \e_1 := (1, 0)^\T & \text{otherwise}
    \end{cases}
\end{align}
For example, $w_{(2)} = u \otimes \e_1 \otimes \e_1 \otimes u$ and $w_{(3)} = u \otimes \e_1 \otimes \e_1 \otimes \e_1 \otimes u \otimes \e_1 \otimes \e_1 \otimes \e_1 \otimes u$. The vector $w_{(D)}$ is a tensor product of $D$ copies of $u$ and of vectors of Hamming weight one, hence it has weight $2^D$. Note that exactly one PCM in the tensor products given in Eqs.~\eqref{PCM_D_X} and~\eqref{PCM_D_Z} is in a position $\ell$ of the form $(i-1)D+i$ and therefore we obtain $H_{\mathrm{SPC}(D)}^x w_{(D)} = H_{\mathrm{SPC}(D)}^z w_{(D)} = 0$. To see that $w_{(D)}$ is not in the image of $(H_{\mathrm{SPC}(D)}^x)^\T$ or of $(H_{\mathrm{SPC}(D)}^z)^\T$ consider the vector 
\begin{align}
\label{v_D}
    v_{(D)} = \bigotimes_{\ell=1}^{D^2} b_\ell
    \qquad
    b_\ell :=
    \begin{cases}
    \e_1 := (1, 0, \underbrace{0 ~ \cdots ~ 0}_{2(s-1)})^\T & \text{if } \text{for } \ell=(i-1)D+i, \; i=1,\dots D \\
    u' \, := (1, 1)^\T & \text{otherwise}
    \end{cases}
\end{align}
Now, note that $H_{\mathrm{SPC}(D)}^x v_{(D)} = H_{\mathrm{SPC}(D)}^z v_{(D)} = 0$ while $ w_{(D)}^\T v_{(D)} = 1$.

\subsection{Meta-check distance for SPC(3, s) codes: $d_M = 3$}

We now show that for a SPC($3,s$) code the meta-check distance is exactly $d_M = w_c = 3$ and therefore it can be used to detect and locate one faulty syndrome. Up to a permutation of the columns, the PCMs for both $X$ and $Z$ checks of the SPC($3,s$) have the form
\begin{align}
  H_{\mathrm{SPC}(3,s)} = 
  \left(
    \begin{array}{ccccc}
	  H_s \o I \o I\\
	  I \o H_s \o I\\ 
	  I \o I \o H_s 
    \end{array}
  \right)
\end{align}
where $H_s = (1 \cdots 1)$ is the PCM associated to a $[8s, 8s-1]$ SPC code. An associated meta-check matrix can be constructed as the following block matrix:
\begin{align}
  M_{\mathrm{SPC}(3,s)} = 
  \left(
    \begin{array}{ccccc ccccc ccccc}
	&&\!\!\!\!0&& & I \o 1 \o H_s & I \o H_s \o 1 \\
	1 \o I \o H_s & &&\!\!\!\!0&& & H_s \o I \o 1 \\
	1 \o H_s \o I & H_s \o 1 \o I & &&\!\!\!\!0&&   
    \end{array}
  \right)   
\end{align}
and one can easily verify that $M_{\mathrm{SPC}(3,s)} H_{\mathrm{SPC}(3,s)} = 0$. The corresponding meta-check code $\mathcal{S}$ has distance three if and only if all weight one errors $e \in \F_2^{192s^2}$ result in distinct nonzero meta-syndrome outcomes $M_{\mathrm{SPC}(3,s)}e \in \F_2^{24s}$. To show it, notice that for any unit vectors $\e_i, \e_j \in \F_2^{8s}$ we have $(I \otimes H_s \otimes 1) (\e_i \otimes \e_j \otimes 1) = \e_i \otimes 1 \otimes 1$ and similarly for other permutations of the tensor components; note that $\e_i \otimes 1 \otimes 1$ is different from zero. It is then possible to determine from the position of the non-zero entries of $M_{\mathrm{SPC}(3,s)}e$ whether the non-zero entry of the error $e$ is located in the first, second or third block of $64s^2$ bits. Supposing, without loss of generality, that the non-zero entry of $e$ is located in one of the first $64s^2$ bits means that $e = 1 \otimes \e_i \otimes \e_j$ for some values $i,j$. Then the two meta-syndrome outcomes $(1 \otimes I \otimes H_s) e =  1 \otimes \e_i \otimes 1$ and $(1 \otimes H_s \otimes I) e =  1 \otimes 1 \otimes \e_j$ contain sufficient information to determine the values $i,j$ and thus locate the position of the syndrome error.

\section{Quaternary belief propagation decoding of stabiliser codes}
\label{app:BP}

We describe an efficient implementation of quaternary BP decoding used in this work. This is a message passing algorithm where information about the likelihood of a Pauli error $\cP=\{I, X, Y, Z\}$ on a qubit is exchanged along the edges of a graph. 

\subsection{Preliminaries and notation}

\renewcommand{\vH}{\mathrm{H}}
\renewcommand{\ve}{\mathrm{E}}
\renewcommand{\vi}{\mathrm{I}}
\renewcommand{\vx}{\mathrm{X}}
\renewcommand{\vy}{\mathrm{Y}}
\renewcommand{\vz}{\mathrm{Z}}
\renewcommand{\vp}{\mathrm{P}}

\paragraph{Stabiliser code representation} 

Let $\cS = \{S_1, \ldots, S_m\}$ be a set of generators of a stabiliser group for a quantum linear code, where each generator is a $n$-qubit Pauli operator, $S_i \in \cP_n$. The set of generators $\cS$ can be specified via a matrix $\vH \in \F_4^{m\times n}$ where each matrix element $\vH_{i,j}$ is in $\F_4 = \{\vi, \vx, \vy, \vz\}$ 
and in one-to-one correspondence to a Pauli operator $\cP =\{I,X,Y,Z\}$. 
The rows of the matrix $\vH$ correspond to different stabilisers while the columns correspond to  different physical qubits. The matrix $\vH$ is sufficient to fully specify a stabiliser code, in contrast with the notation used in the rest of the paper where two binary matrices ($H^x$ and $H^z$) are needed to define a CSS code.

\paragraph{Commutation and anti-commutation relations}

We define an inner product $\inner{\cdot}{\cdot} : \F_4 \times \F_4 \rightarrow \F_2$ as
\begin{center}    
\begin{tabular}{|c|cccc|}
     \hline
     $\inner{\cdot}{\cdot}$ & 
     $\vi$&$\vx$ & $\vy$&$\vz$\\ 
     \hline
     $\vi$& 0&0&0&0 \\
     $\vx$& 0&0&1&1 \\
     $\vy$& 0&1&0&1 \\
     $\vz$& 0&1&1&0 \\
     \hline
\end{tabular}
\end{center}
i.e., the inner product $\langle \cdot,\cdot\rangle$ is equal to $0$ if the corresponding Pauli matrices commute and is equal to 1 if the Pauli matrices anti-commute. The commutativity of the stabiliser generators then translates to
\begin{align}
    \sum_{j=1}^n 
    \inner{\vH_{i,j}}{\vH_{k,j}} = 0 \mod 2
    \qquad \forall\, i,k \in\{1,\ldots,m\}
\end{align}
which generalises the CSS commutativity condition in Eq.~\eqref{eq:CSS}. Similarly, given a $n$-qubit Pauli error $E \in \cP_n$ having associated a vector $\ve \in \F_4^n$, the corresponding syndrome outcomes $s \in \F_2^m$ are obtained as
\begin{align}\label{eq:si}
    s_i = \sum_{j=1}^{n} \inner{\vH_{i,j}}{\ve_j} \mod 2\,.
\end{align}
Let the set $\cZ_{i,j}$  and $ (\overline{\cZ}_{i,j})$ be 
\begin{align}
    \cZ_{i,j} := 
    \big\{\,\vp \in \F_4\, \big\vert \
    \inner{\vH_{i,j}}{\vp} = 0 \big\}
    \qquad
    \overline{\cZ}_{i,j} := \F_4 \setminus \cZ_{i,j} \,.
\end{align}
In other words, $\cZ_{i,j}$ ($\overline{\cZ}_{i,j}$) corresponds to the set of Pauli operators that stabiliser $S_i$ commutes with (anti-commutes with) at qubit $j$.

\paragraph{Quantum Tanner graph representation}

\begin{figure}[t]
    \centering
    \begin{tabular}{|c|}
    \hline \\[-2mm]
    \textbf{CN update rule in $\F_2$} \\[1mm]
    \includegraphics[width=9cm]{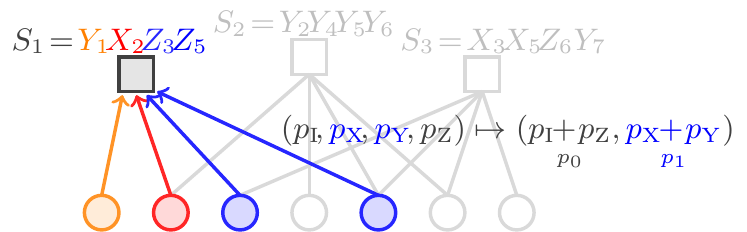} \\[1mm]
    \hline \\[-2mm]
    \textbf{VN update rule in $\F_4$} \\[1mm]
    \includegraphics[width=9cm]{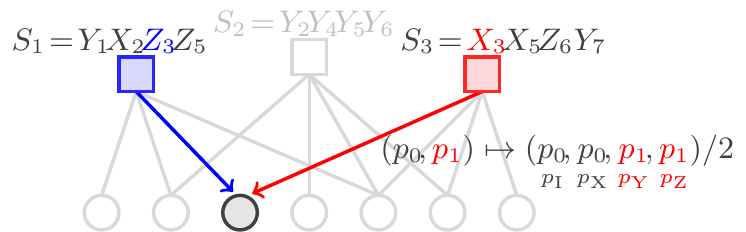} \\[1mm]
    \hline
    \end{tabular}
    \caption{Example binary/quaternary inter-conversion in message passing for a quantum Tanner graph having 7 VNs (qubits) and 3 CNs (stabiliser generators). Edges transmitting messages to one CN $\cn_j$ (top) and to one VN $\vn_i$ (bottom) are highlighted and the edges are labelled (coloured) according to the Pauli operator measured by the stabiliser $S_i$ on qubit $j$. }
    \label{fig:updates}
\end{figure}

The code's stabiliser matrix $\vH$ can be represented as a quantum Tanner graph, i.e., a labeled bipartite graph having two distinct sets of vertices, called \textit{variable nodes} (VNs) and \textit{check nodes} (CNs), as well as a set of $\F_4$-labeled edges connecting vertices of different types. VNs are associated to physical qubits and CNs to stabiliser generators. An edge connects a VN $\vn_j$ to a CN $\cn_i$ if and only if $\vH_{i,j} \in \F_4 \setminus \{\vi\}$. The edge label corresponds to $\vH_{i,j}$ (for $\vH_{i,j} \neq \vi$). The set of neighbours of a VN $\vn_j$ (or of a CN $\cn_i$) is denoted by $\cN(\vn_j)$ ($\cN(\cn_i)$). See Figure~\ref{fig:updates} for a graphical representation.

\paragraph{Message passing}
Decoding can be cast as passing messages between VNs and CNs along the edges of the Tanner graph. In a successive manner, all CNs send messages to all the neighbouring VNs, and all VNs exchange messages with all the neighbouring CNs until a stopping criterion is reached. CNs and VNs process incoming messages according to certain rules. Messages are (for instance) quaternary probability vectors representing an estimate of the probability of a given value of $\ve_j$. 
Observe that  the inner product in Eq.~\eqref{eq:si} yields the same binary value for two distinct values of $\ve_j$, i.e., $\inner{\vH_{i,j}}{\ve_j}$ is zero for $\ve_j \in \cZ_{i,j}$ and one for $\ve_j \in \overline{\cZ}_{i,j}$. Hence, we can add up the respective probabilities and apply the CN processing on a binary probability vector, rather than on a quaternary probability vector.

\paragraph{Log-domain decoding}
For numerical stability message passing decoders are often implemented using ratios of logarithmic probabilities, referred to as log-likelihood ratios (LLR). For probability estimates $p_0$ and $p_1 = 1-p_0$ of binary outcomes we define the associated LLR as
\begin{align}
    L = 
    \log \frac{p_0}
            {p_1} .
\end{align} 
For quaternary outcomes with probabilities $p_\vi, p_\vx, p_\vy, p_\vz$ we define an LLR vector as
\begin{align}
    \cL = 
    \left(0,
    \log \frac{p_\vx}{p_\vi},
    \log \frac{p_\vy}{p_\vi},
    \log \frac{p_\vz}{p_\vi} \right).
\end{align} 

\paragraph{Conversions between binary and quaternary message passing}

Consider a stabiliser $S_i$ and a qubit $j$.  
Quaternary messages from $\vn_j$ to $\cn_i$ are converted to binary messages by computing $p_1$ as the sum of the probabilities associated to operators in $\overline{\cZ}_{i,j}$ and then $p_0 = 1-p_1$. Binary messages from $\cn_i$ to $\vn_j$ are converted to quaternary messages by equally splitting the probability $p_1$ among the operators in $\overline{\cZ}_{i,j}$ and equally splitting $p_0$ among the other two Pauli operators (i.e., those in $\cZ_{i,j}$). See Figure~\ref{fig:updates} for an example. For conversion of quaternary to binary messages, we will make use of the soft-max operator ($\maxs$) which is defined as
\begin{align}
    \maxs(a,b) 
    := &\, \log(\exp(a)+\exp(b))\\
     = &\, \max(a,b)+\log\left(1+\exp\left(-|a-b|\right)\right).
\end{align}
The first expression makes it obvious that $\maxs$ is symmetric and associative, the second expression is the one that shall be used for a numerically stable implementation. 

\subsection{Update rules}
\label{sec:rule_description}

Let us focus on iteration number $t$. An iteration consists in messages sent from CNs to VNs followed by messages sent from VNs to CNs.

Let the message received by CN $\cn_i$ from the VN $\vn_j$ be $L_{\vn_j\rightarrow \cn_i}^{(t-1)}$. The message that the CN $\cn_i$ sends to VN $\vn_j$, $L_{\cn_i\rightarrow \vn_j}^{(t)}$, is obtained by using the standard update rule for binary message passing decoders (embedding the syndrome value $s_i$):
\begin{align}
    L_{\cn_i\rightarrow \vn_j}^{(t)} 
    & = 
    (-1)^{s_i} \, 2 \tanh^{-1}
    \left(\prod_{j':\, \vn_{j'} \in \cN(\cn_i)\setminus \vn_j} \tanh \left( \frac{1}{2} L_{\vn_{j'}\rightarrow \cn_i}^{(t-1)}\right)\right). 
    \label{eq:CN_update}
\end{align}
\begin{remark}
There are a manifold of approximations of the \ac{CN} update rule in the literature which permit a hardware-friendly implementation. 
\end{remark}

The LLR $L_{\cn_i\rightarrow \vn_j}^{(t)}$ is converted into a quaternary LLR $l_{\cn_i\rightarrow \vn_j}^{(t)}$ for further processing at the VNs:
\begin{align}
   \cL_{\cn_i\rightarrow \vn_j}^{(t,\vp)} 
   & = 
   \begin{cases}
    0 & \text{for } \vp \in \cZ_{i,j} \\
    - L_{\cn_i\rightarrow \vn_j}^{(t)} & \text{for } \vp \in \overline{\cZ}_{i,j}
    \label{eq:LPR_vector_VN}
\end{cases}
\end{align}
where $\vp \in \F_4$ indexes the values of the message. 
The VN update operation is the standard one for quaternary message passing and yields the error probability estimates 
\begin{align}
    \cL_{\vn_{j}\rightarrow \cn_i}^{(t)}
    & = 
    \overline{\cL}_j + 
    \sum_{i':\,\cn_{i'} \in \cN(\vn_j)\setminus \cn_{i}} \cL_{\cn_{i'}\rightarrow \vn_j}^{(t)}
    \label{eq:vn_operation}
\end{align}
where $\overline{\cL}_j$ is the channel message (a.k.a., \textit{prior} LLR) and is a $\vp$-indexed vector which depends on the channel model. For the depolarising noise in Eq.~\eqref{eq:depolarising} we have
\begin{align}
   \overline{\cL}_j^{(\vp)} 
   & =
   \begin{cases}
    0 & \text{for } \vp = \vi \\
    \log\left(\frac{\epsilon}{3 (1-\epsilon)}\right) & 
    \text{for } \vp\in \F_4 \setminus \{\vi\}
    \end{cases} .
\end{align}

\begin{remark}
The evaluation of Eq.~\eqref{eq:vn_operation} can be simplified by avoiding to construct a quaternary LLR vector $\cL^{(t)}_{\cn_{i'}\rightarrow \vn_j}$. It is sufficient to add its two non-zero elements to the respective elements of $\cL^{(t)}_{\vn_{j}\rightarrow \cn_i}$ in \eqref{eq:vn_operation}.
\end{remark}

Finally, we convert $\cL_{\vn_{j}\rightarrow \cn_i}^{(t)}$ to a scalar LLR $L^{(t)}_{\vn_{j}\rightarrow \cn_i}$ that is given as input the respective CN,
\begin{align}
    L^{(t)}_{\vn_{j}\rightarrow \cn_i} & = 
    \maxs_{\vp\in \cZ_{i,j}}(\cL_{\vn_{j}\rightarrow \cn_i}^{(t,\vp)}) - 
    \maxs_{\vp\in \overline{\cZ}_{i,j}}(\cL_{\vn_{j}\rightarrow \cn_i}^{(t,\vp)}) .
    \label{eq:VN_operation_binary}
\end{align}

\subsection{Initialization and final decision}

We give the message from VN $\vn_j$ to CN $\cn_i$ at the beginning of the very first decoding iteration. We have 
\begin{align}
    \cL^{(0)}_{\vn_{j}\rightarrow \cn_i} &= \overline{\cL}_{j} 
\end{align}
and from \eqref{eq:VN_operation_binary} we obtain
\begin{align}
    L^{(0)}_{\vn_{j}\rightarrow \cn_i} &= \log\left(\frac{3-2 \epsilon}{2 \epsilon} \right).
\end{align}

A hard decision $\hat{\ve}_j$ on the value of $\ve_j$ is made according to
\begin{align}
    \hat{\ve}_j & = 
    \argmax_{}\left(\overline{\cL}_j + \sum_{i: \cn_{i} \in \cN(\vn_j)} \cL_{\cn_{i}\rightarrow \vn_j}\right).
    \label{eq:hard_decision}
\end{align}
Recall that for convenience the elements of quaternary LLR are indexed by $\textup{P}\in \F_4$. The decoder returns this hard decision after a maximum number of decoding iterations is reached or when the estimated error vector $\hat{\ve}$ satisfies all parity-checks.

\subsection{Inclusion of meta-checks}

The Tanner graph can be expanded to include readout errors and meta-checks. In particular, a new VN is added for each CN, which represents the classical readout error channel. These new VN are therefore labelled as ``binary'' and will process the received messages differently from the ``quaternary'' representing the Pauli errors on the qubits. Furthermore, these new variable nodes are connected to a new set of CNs, corresponding to the meta-checks. The extended Tanner graph is associated to a labelled adjacency matrix of the form
\begin{align}
    \widetilde{\textup{H}} = 
    \left(\begin{array}{c|c}
    ~\textup{H}~ &  ~I~ \\ \hline
    0   &   M  \\     
    \end{array} \right)
\end{align}
where the leftmost columns (comprising $\textup{H}$ and $0$) have entries in $\F_4$ and the rightmost columns (comprising $I$ and $M$) have entries in $\F_2$. The BP decoder then becomes hybrid, where either quaternary or binary messages are passed, depending on the VN type ($\F_4$-type or $\F_2$-type).

The BP update rules are modified as follows. The CN to VN messages are always binary and are directly given by Eq.~\eqref{eq:CN_update}, for the extended set of VN, CN and neighbourhoods $\cN(\cn_i)$. The channel message for each new variable node $\vn_j$ is given by $\overline{L}_j = \log \frac{p}{1-p}$, where $p$ is the probability of flipping the readout value. Thus, the VN to CN are computed in two different ways, depending on the VN type:
\begin{align}
    L^{(t)}_{\vn_{j}\rightarrow \cn_i} & = 
    \begin{cases}
    \maxs_{\vp\in \cZ_{i,j}}(\cL_{\vn_{j}\rightarrow \cn_i}^{(t,\vp)}) - 
    \maxs_{\vp\in \overline{\cZ}_{i,j}}(\cL_{\vn_{j}\rightarrow \cn_i}^{(t,\vp)})
    \label{eq:VN_operation_binary_meta_checks}
    & 
    \F_4\text{-type VN}\\
    \overline{L}_j + 
    \sum_{i':\,\cn_{i'} \in \cN(\vn_j)\setminus \cn_{i}} L_{\cn_{i'}\rightarrow \vn_j}^{(t)}
    & 
    \F_2\text{-type VN}\\
    \end{cases}
\end{align}
with the notation given in Section~\ref{sec:rule_description}. The hard decision for the data-qubit errors, Eq.~\eqref{eq:hard_decision}, is unmodified, while the estimate $\hat{e}_s$ of the readout errors $e_s$ is obtained via the hard decision
\begin{align}
    (\hat{e}_s)_j & = 
    \argmax_{}\left(\overline{L}_j + \sum_{i: \cn_{i} \in \cN(\vn_j)} L_{\cn_{i}\rightarrow \vn_j}\right) .
\end{align}